# Risk-aware Flexible Resource Utilization in an Unbalanced Three-Phase Distribution Network using SDP-based Distributionally Robust Optimal Power Flow

Zelong Lu, *Student Member, IEEE,* Jianxue Wang, *Senior Member, IEEE,* Mohammad Shahidehpour, *Fellow, IEEE,* Linquan Bai, *Senior Member, IEEE*, Zuyi Li, *Senior Member, IEEE,* Lei Yan, Xianlong Chen

*Abstract*—The variability caused by the proliferation of distributed energy resources (DERs) and the significant growth in unbalanced three-phase loads pose unprecedented challenges to distribution network operations. This paper focuses on how a distribution system operator (DSO), taking over the distribution grid and market operations, would develop a risk-aware flexibility market to mitigate uncertainties in an unbalanced three-phase power distribution network. First, a distributionally robust chance constraint (DRCC) method is devised to solve the unbalanced three-phase optimal power flow using a semidefinite programming (SDP) model. The DSO can apply the proposed solution to jointly clear energy and flexibility markets. Then, the DRCC model accuracy is improved by an information-sharing mechanism characterized by spatially-correlated uncertainties in the distribution grid. Further, a novel system-wide response function is derived to make the DRCC model tractable. Using the duality theory, the paper further investigates the physical composition of the DSO's cleared flexibility prices to guide the unbalanced distribution network operation. Finally, the effectiveness of the risk-aware flexibility market is verified in a modified three-phase IEEE 34-node system. Results demonstrate that the flexibility market can quantify the impact of spatially correlated uncertainties and facilitate the utilization of flexible resources to mitigate uncertainties across the network.

*Index Terms*—Unbalanced three-phase distribution system, DRCC optimal power flow, risk-aware flexible resources.

## Nomenclature

### A. Indices and Sets

| | |
|---|---|
| $t, \mathrm{T}$ | Index and set of time periods. |
| $i, \varphi/\phi$ | Indices of nodes and phases. |
| $k$ | Indices for uncertainty sources. |
| $l/(i,j)$ | Indices for branch lines. |
| $\Omega_b, \Omega_l$ | Set of nodes and lines. |
| $\Omega_b^+$ | Set of nodes except the substation node. |
| $\Omega_g, \Omega_s$ | Set of nodes with generators/ storages. |

### B. Main Variables

| | |
|---|---|
| $g_{i,\varphi,t}, q_{i,\varphi,t}^g$ | Active/reactive output of generator at phase $\varphi$ of bus $i$ at time $t$. |
| $\mathrm{ch}_{i,t}^\varphi/\mathrm{dis}_{i,t}^\varphi$ | Charging and discharging power of storage at phase $\varphi$ of bus $i$ at period $t$. |
| $q_{i,\varphi,t}^s$ | Reactive output of storage located at phase $\varphi$ of bus $i$ at period $t$. |
| $R_i^{\mathrm{up}}/R_i^{\mathrm{dn}}$ | Up/Down reserve by bus $i$. |
| $P_{i,\varphi,t}, Q_{i,\varphi,t}$ | Active/Reactive power injection at phase $\varphi$ of bus $i$ at period $t$. |
| $VDI_{i,t}$ | Voltage deviation index for bus $i$ at time $t$. |
| $P_{i,j,t}^\varphi, Q_{i,j,t}^\varphi$ | Real/Reactive power flow at phase $\varphi$ of line $(i,j)$ at time $t$. |
| $SOC_{i,t}$ | State of charge for the storage at bus $i$ at time $t$. |
| $\boldsymbol{V}_t$ | Complex nodal voltage vector at time $t$. |
| $\boldsymbol{X}_t$ | Auxiliary variables $\boldsymbol{X}_t = [\Re(\boldsymbol{V}_t^T), \Im(\boldsymbol{V}_t^T)]^T$, where $\boldsymbol{V}_t = [V_{1,t}^a, V_{1,t}^b, V_{1,t}^c, \ldots, V_{N,t}^a, V_{N,t}^b, V_{N,t}^c]^T$ |
| $\boldsymbol{W}_t$ | Symmetric matrix $\boldsymbol{W}_t = \boldsymbol{X}_t \boldsymbol{X}_t^T$ |
| $\boldsymbol{S}_t$ | Power matrix $\boldsymbol{S}_t = [\boldsymbol{P}_t; \boldsymbol{Q}_t]$ $= \begin{bmatrix} P_{1,t}^a, P_{1,t}^b, P_{1,t}^c, \ldots, P_{N,t}^a, P_{N,t}^b, P_{N,t}^c \\ Q_{1,t}^a, Q_{1,t}^b, Q_{1,t}^c, \ldots, Q_{N,t}^a, Q_{N,t}^b, Q_{N,t}^c \end{bmatrix}^T$ |
| $\hat{S}_{i,j,t}$ | Apparent power of line $(i,j)$ flow at time $t$. |
| $\beta_{i,k,t}^\varphi$ | Proportion of uncertainty source $\xi_k$ which will be balanced by phase $\varphi$ of bus $i$ at time $t$. |
| $\boldsymbol{\xi}_t$ | Vector of uncertainties at time $t$ |
| $C_{i,t}^R$ | Cost for flexibility at bus $i$ at time $t$. |
| $E_{k,t}^\xi$ | Uncertainty payment for source $k$ at time $t$. |

Other Greek letters are used to represent the dual multipliers.

### C. Main Parameters

| | |
|---|---|
| $RU, RD$ | Ramp up/down rate. |
| $PF_i^{min/max}$ | Minimum/maximum power factor for bus $i$ generator. |
| $\eta_s$ | Efficiency factor of storage. |
| $\boldsymbol{Y}$ | System admittance matrix. |
| $S_{i,j}^{max}$ | Maximum apparent power limit for line $(i,j)$ |
| $d_{i,\varphi,t}$ | Load forecast at phase $\varphi$ of bus $i$ at time $t$. |
| $e_{i,\varphi,t}$ | Renewable energy forecast at phase $\varphi$ of bus $i$ at time $t$. |
| $\phi_{i,\varphi}^d$ | Load power factor. |
| $V_i^{min}, V_i^{max}$ | Minimum/maximum voltage magnitude. |
| $\boldsymbol{\Gamma}$ | Covariance matrix of uncertainty. |
| $\boldsymbol{\Gamma}^{1/2}$ | Decomposed covariance matrix satisfying: $(\boldsymbol{\Gamma}^{1/2})^T \boldsymbol{\Gamma}^{1/2} = \boldsymbol{\Gamma}$ |
| $\epsilon_R, \epsilon_v, \epsilon_f$ | Confidence level (probability) of reserve, voltage, line flow constraint violations. |
| $\boldsymbol{\rho}$ | Matrixes of correlation coefficients. |
| $\underline{\beta}$ | Minimum participation factor for flexibility reserve requirement. |

### D. Main Abbreviations and operator

| | |
|---|---|
| $\Re(), \Im()$ | Real and imaginary parts. |
| $Tr\{\}, ()^T$ | Trace/transpose. |
| $E()$ | Expected value. |
| $Stdev[]$ | Standard deviation |
| $cov(\cdot)$ | Covariance of the vector |
| ⊙ | Element-wise product. |
| $\Phi_N$ | Joint probability density function of N standard normal distribution functions. |
| $\Phi^{-1}$ | Inverse of univariate standard normal CDF distribution. |
| CDF | Cumulative distribution function |
| SDP | Semidefinite programming. |
| DRCC | Distributionally robust chance constraints. |
| DERs | Distributed energy resources. |

## I. INTRODUCTION

POWER distribution network is envisioned to incorporate a higher penetration level of renewables, plug-in three-phase electrical loads, and plug-and-play distributed energy storage system (ESS). With a higher penetration of distributed energy resources (DERs) and intermittent loads, spatial and temporal distribution of power generation to supply the load demand in three-phase unbalanced distribution networks is expected to be extremely uncertain [1]-[3]. Generally, more flexibility is required to maintain the energy balance, avoid voltage violations, and mitigate distribution flow congestions.

Extensive studies have explored the value of flexible resources in distribution networks [1]-[3]. The North American Electric Reliability Corporation (NERC) recognizes the flexibility as an ability to mitigate fluctuations in renewable energies or loads by using controllable resources in a system [4]. One of the essential roles of flexibility is the provision of reserve, which is generally regarded as an ancillary service [5]. However, the provision of system-wide flexibility may face a deliverability issue restricted by network constraints [6]. On the other hand, ancillary services may face significant shortages due to rising uncertainties in power systems [7]. Two categories of probability-related methods are explored to analyze the influence of uncertainties on the network operation.

(i) *Probabilistic optimal power flow (POPF) method*: POPF determines the optimal network operation strategy considering inherent uncertainties by analyzing probabilistic distribution characteristics of system-state variables [8]-[10]. Probability distributions of system-state variables can be obtained directly by probability density functions (PDF) or cumulative distribution functions (CDF) to assess the system performance [8], [10]. Nonlinear complementarity problem functions have been introduced in [9] to derive nonlinear analytical expressions for optimal solutions. PDF or CDF of the solution can be reformulated using various approximation expansions such as Edgeworth, Gram-Charlier, and Cornish-Fisher expansions [11], [12]. However, while POPF accurately captures the impact of uncertainties on the system state, it often requires computationally intensive calculations [8], [10].

(ii) *Distributionally robust chance constraints-based optimal power flow (DRCC-OPF) method*: To overcome the challenges posed by the heavy computational burden of the POPF method, the DRCC-OPF method, as another probabilistic model, has attracted widespread attention. It makes a tradeoff between cost-efficiency and system reliability by considering maximum probability violations for operational constraints. DRCC-OPF models uncertainty using an ambiguity set, a collection of potential distributions with uncertain characteristics [13]. The distance-based DRCC methods proposed in [14]-[16] capture the proximity of adjacent distributions using appropriate probability distance functions like Prokhorov metric [14], Kullback-Leibler divergence [15], and Wasserstein distance [16]. While these methods provide detailed probability distribution characterization, they come with an increased computational burden and challenges in embedding uncertainty dependencies into the OPF problem. To mitigate these limitations, moment-based ambiguity sets are employed in the DRCC-OPF problem in [17]. These sets use moment information, such as mean and covariance, to describe the probability distributions of uncertainties. The ambiguity set comprises distributions that share the same mean and covariance matrix estimated from empirical data.

The Distribution System Operator (DSO), responsible for grid and market operations, places significant importance on making robust market decisions given the wide range of probability distribution information associated with uncertainties. While the DRCC-OPF method demonstrates efficiency in evaluating network operation under diverse uncertainties, it still faces three primary challenges:

1) *How can we assess the influence of spatially correlated uncertainties on distribution network operation?* Numerous studies have pointed out that there are spatial correlations between uncertain variables in the distribution network, including correlations between loads under different spatial distributions [18], correlations between DERs located at different nodes [19], and correlations between DERs and loads [6]. Ref. [20] develops a combined Gaussian copula model with Spearman coefficients to analyze the rank correlations among diverse uncertainties that would follow any probability distribution. Although uncertainty correlation analyses in distribution networks have attracted widespread attention, the issue that needs to be resolved in how to precisely assess the impact of spatially correlated uncertainty on the operation safety of three-phase distribution networks.

Incorporating the correlations among uncertainties in the distribution systems into the DSO's decision model can enhance robustness and reduce the requirement for flexibility reserves [6], [17]. For the unbalanced three-phase distribution network, the rank-relaxed semidefinite program (SDP) is considered as a valid solution to solve the OPF model in unbalanced three-phase distribution networks [21]. [22] verifies the tightness of the SDP relaxation in the presence of practical angle constraints and real power lower bounds in radial systems. Hence, it is promising to incorporate the probability information of spatially correlated uncertainties into the SDP-based ACOPF in the three-phase distribution network.

2) *How will DSO estimate the flexibility requirement to improve the operation of three-phase unbalanced distribution network operation?* Although certain studies focused on correlated uncertainties in the OPF model, it remains a challenge to tie the spatially correlated uncertainties with flexibility reserve requirements. In nonlinear ACOPF, the system-state variables (such as node voltage, active and reactive power flows in the line) are often implicitly relative to the changes in node power injection uncertainty. To quantify the flexibility requirement of distribution network operation, it is essential to derive an explicit mapping relationship between the uncertainties and the change of system-state variables from the original ACOPF [23], [24]. The affine policy is regarded as an efficient tool expansion to derive the expressions for the system-state variables affected by uncertainties [25]. The traditional affine policy can recast the nonlinear node power balance equations in ACOPF through a Jacobian matrix [23], [24]. In the existing research on ACOPF in distribution networks, there are two common methods to obtain the Jacobian matrix. One is to approximate the nonlinear ACOPF with linear LinDistFlow [26]. Further, the power transfer distribution factor (PTDF) is used to map the change of nodal load to the change of edge power flow [27]. The other method is the algorithmic differentiation method [28], which can effectively calculate the Jacobian matrix corresponding to various operating states. However, for the first kind, we cannot neglect the influence of unbalanced power injections and corresponding reactive power flows in three-phrase distribution networks. Hence, the accuracy of the traditional PTDF method is unaccepted. For the second kind, as shown in [27], [28], this method requires recalculation at all given operating points to obtain the Jacobian matrix, which will cause a considerable computational burden. Hence, how to efficiently assess the requirement of flexibility reserves in the unbalanced distribution network is still an open question.

*3) How do we design a risk-aware distribution network flexibility market for mitigating the operation risk arising from diverse uncertainties?* Considering market operation perspectives, the cost of ancillary services is usually passed on to customers using fixed tariffs. In the PJM market [29], a reserve bill is allocated to each load serving entity (LSE) according to its load share. Under such a tariff, LSE customers, instead of the owner of variable sources, pay for the flexibility required to mitigate the variability from DER generation and load demands. Such practiced do not provide an effective incentive for local (i.e., located in the distribution system) flexibility resources, since variable DERs would have no incentives to manage their risk posed on distribution system operations [30].

The concept of distribution-level flexibility markets has emerged recently [31]. A large body of work has recently focused on local flexibility markets, including bidding strategies for flexibility offers [32], and efforts to account for network constraints [33], uncertainty [34]. Mechanism design is used to design a fair market for energy and flexibility at the distribution level in [35]. In [30], the authors move even a step further and present an approach toward understanding distribution locational marginal prices by decomposing the distribution locational marginal price (DLMP) for energy into terms relating to power at the root node, to real power losses, to reactive power losses, to voltage constraints, and to line limits. However, DLMP is hindered by its inability to clearly capture the DER stochasticity [23]. [36] proposes a SDP-based DRCC-OPF model to manage the risk of operational limits violations which are caused by uncertain renewable generation. It is a promising solution for assessing the effect of uncertainties on the single-phase distribution network operation. However, the unbalanced distribution network features are neglected and all risk costs are shifted to end-users. Hence, we derive risk-aware prices on both sides of the flexibility market to quantify the cost of mitigating the uncertainties and reward flexible resources in an unbalanced three-phase distribution network.

Motivated by the above challenges, this paper proposes a risk-aware distribution-level flexibility market clearing scheme for valuing local flexible resources that can cope with spatially correlated uncertainties. Through the clearing of flexibility market, DSO would quantify the cost of mitigating the risk posed by uncertainties and reward flexible resources in an unbalanced three-phase distribution network. First, DSO jointly clears energy and flexibility markets using a SDP-based three-phase ACOPF solution. To further model the impact of spatially correlated uncertainties on flexibility market clearing, a DRCC framework is included in the probabilistic three-phase ACOPF model with an information-sharing mechanism that assesses the correlation amongst stochastic variables. Finally, the cleared risk-aware flexibility prices are extended with a transparent decomposition for DSO to value both the benefit of flexible resources and the risk posed by uncertainties on system operation in an unbalanced three-phase distribution network.

The contributions of this paper are threefold:

i) We internalize spatially correlated uncertainties and the system-wide risk tolerance level through information sharing in a DRCC based unbalanced three-phase distribution ACOPF model. The derivation can be applied to diverse uncertainty factors with arbitrary distributions which are coupled spatially. The DRCC model can mitigate the total system-level reserve requirement by accurately characterizing network uncertainties with shared information.

ii) System-wide response functions are derived from sensitivity matrices of an unbalanced three-phase OPF model. The functions can analytically characterize the spatial-temporal transfer relationship of uncertainties and operation flexibilities in a power distribution system. Using the system-wide response functions, the proposed DRCC model will be reformulated as a tractable SDP problem.

iii) A DSO's risk-aware flexibility market clearing scheme is derived from the probabilistic three-phase ACOPF solution. The cleared flexibility prices are further decomposed into four physical parts including energy, volt/var, and active/reactive power flows. This decomposition is used for guiding flexible resources to mitigate the impact of uncertainties on energy balance, voltage security, line losses, and network congestion.

The rest of this article is organized as follows. Section II performs an SDP relaxation method for the deterministic unbalanced three-phase ACOPF. Then the SDP-based DRCC-OPF solution is used to clear the risk-aware flexibility market in Section III. Section IV analyzes the cleared flexibility prices. Section V validates the effectiveness of the proposed pricing framework. The conclusion is performed in Section VI.

## II. SDP Relaxation for ACOPF in an Unbalanced Three-Phase Power Distribution Network

This section shows the SDP relaxation of the three-phase unbalanced distribution OPF model. Further, it constructs sensitivity matrices to represent nodal uncertain deviations as a function of system-state variables.

*A. Preliminaries and Matrix Transformation*

In this section, we present the three-phrase unbalanced power flow constraints as a function of auxiliary variables and matrices [37]. Auxiliary variables $Y_i^\varphi, \overline{Y}_i^\varphi, M_i^\varphi, \Phi_{i,j}^\phi$, and $\overline{\Phi}_{i,j}^\phi$ are defined in Appendix A.

A symmetric matrix $W_t$ is defined to exhibit the coupling between system variables [37]:
$$W_t = X_t X_t^T \quad (1a)$$
where $X_t$ is based on the complex voltage variables,
$$X_t = [\Re(V_t^T), \Im(V_t^T)]^T \quad (1b)$$
$$V_t = [V_{1,t}^a, V_{1,t}^b, V_{1,t}^c, \ldots, V_{N,t}^a, V_{N,t}^b, V_{N,t}^c]^T \quad (1c)$$

Accordingly,
$$P_{i,\varphi,t} = Tr\{Y_i^\varphi W_t\}, \ i \in \Omega_b : \beta_{i,\varphi,t}^P \quad (2a)$$
$$Q_{i,\varphi,t} = Tr\{\overline{Y}_i^\varphi W_t\}, \ i \in \Omega_b : \beta_{i,\varphi,t}^Q \quad (2b)$$
$$|V_{i,\varphi,t}|^2 = Tr\{M_i^\varphi W_t\}, \ i \in \Omega_b : \beta_{i,\varphi,t}^V \quad (2c)$$
$$P_{i,j,t}^\varphi = \sum_\phi Tr\{\Phi_{i,j}^{\varphi\phi} \cdot W_t\}, \ (i,j) \in \Omega_l : \beta_{i,j,\varphi,t}^P \quad (2d)$$
$$Q_{i,j,t}^\varphi = \sum_\phi Tr\{\overline{\Phi}_{i,j}^{\varphi\phi} \cdot W_t\}, \ (i,j) \in \Omega_l : \beta_{i,j,\varphi,t}^Q \quad (2e)$$
where the dual multiplier of each constraint is defined.

*B. Unbalanced Three-Phase ACOPF for Distribution Network*

The objective function in the three-phase ACOPF problem is formulated as a polynomial expression aimed at minimizing the overall cost associated with supplying energy and flexibility reserves in an unbalanced distribution network.
$$C_t^{sub} + \sum_{i,t} C_{i,t}^{DER} + \sum_i \left(C_{i,R}^{up}(R_i^{up}) + C_{i,R}^{dn}(R_i^{dn})\right) \quad (3)$$
where the first term is the energy cost from upstream grid which includes active/reactive power purchase cost. The second term denotes the total cost for dispatching DERs, and the last two terms refer to the cost of purchasing up/down reserves from local flexible resources. The cost terms are stated as quadratic or piecewise-linear functions.

The following are respective constraints with their Lagrange multipliers.

➤ Generator and ESS operational limits:
$$A^g \cdot p^g + B^g \cdot q^g + C^g \cdot R^{up,g} + D^g \cdot R^{dn,g} \leq f^g : v^g \quad (4a)$$
$$A^s \cdot p^s + B^s \cdot q^s + C^s \cdot R^{up,s} + D^s \cdot R^{dn,s} \leq f^s : \mu^g \quad (4b)$$
where the decision variables for active/reactive power output, up/down generator reserves, and ESSs are denoted as $p^{g/s}$, $q^{g/s}$, $R^{up,g/s}$ and $R^{dn,g/s}$, respectively. $A^{g/s}$, $B^{g/s}$, $C^{g/s}$, and $D^{g/s}$ are corresponding coefficient matrices. The detailed modeling is provided in Appendix B.

➤ Power flow limits:
$$P_{i,j,t} = \sum_\varphi P_{i,j,t}^\varphi, Q_{i,j,t} = \sum_\varphi Q_{i,j,t}^\varphi, (i,j) \in \Omega_l : \eta_{i,j,t}^p, \eta_{i,j,t}^q \quad (5a)$$
$$(P_{i,j,t})^2 + (Q_{i,j,t})^2 = \hat{S}_{i,j,t}, (i,j) \in \Omega_l : \eta_{i,j,t}^l \quad (5b)$$
$$0 \leq \hat{S}_{i,j,t} \leq (S_{i,j}^{max})^2, (i,j) \in \Omega_l : \underline{\eta}_{i,j,t}^l, \overline{\eta}_{i,j,t}^l \quad (5c)$$
$$(V_i^{min})^2 \leq Tr\{M_i^\varphi W_t\} \leq (V_i^{max})^2 : \underline{\eta}_{i,j,t}^V, \overline{\eta}_{i,j,t}^V \quad (5d)$$
$$P_{i,\varphi,t} = p_{i,\varphi,t}^g + dis_{i,\varphi,t} - ch_{i,\varphi,t}$$
$$\quad -d_{i,\varphi,t} + e_{i,\varphi,t}, i \in \Omega_b^+ : \lambda_{i,\varphi,t}^P \quad (5e)$$
$$Q_{i,\varphi,t} = q_{i,\varphi,t}^g + q_{i,\varphi,t}^s - d_{i,\varphi,t} \cdot \phi_{i,\varphi}^d, i \in \Omega_b^+ : \lambda_{i,\varphi,t}^Q \quad (5f)$$
$$\sum_\varphi P_{1,\varphi,t} = P_{0,t} : \lambda_{0,t}^P \quad (5g)$$
$$\sum_\varphi Q_{1,\varphi,t} = Q_{0,t} : \lambda_{0,t}^Q \quad (5h)$$
$$VDI_{i,t} = \max_\varphi |V_{i,\varphi,t}|^2 - \min_\phi |V_{i,\phi,t}|^2 \leq \varepsilon_v,$$
$$\forall i \in \Omega_b, t \in T : \lambda_{i,t}^V \quad (5i)$$
$$(1a), (2a)-(2e) \quad (5j)$$
where (5a)-(5c) denote power flow constraints. (5d) is the nodal three-phase voltage magnitude limit, and (5e)-(5h) illustrate power injection constraints for active/reactive power balance. The right-hand side variables of (5g)-(5h) represent the injections from electricity and ancillary service markets. Eq. (5i) ensures that the maximum unbalanced square of voltage deviation does not exceed the limit [38], where $VDI_{i,t}$ denotes the voltage deviation index and $\varphi/\phi$ denote phases.

*C. SDP-Relaxation of the Three-Phase ACOPF Model*

The use of nonlinear constraints (1a), (5b), (5i), makes it difficult to solve the original three-phase ACOPF. Herein, an SDP relaxation is developed to make the problem tractable.
$$W_t \succcurlyeq 0 : \eta_t^W \quad (6a)$$
$$\begin{bmatrix} \hat{S}_{i,j,t} & \sum_\varphi Tr\{\Phi_{i,j}^\varphi \cdot W_t\} & \sum_\varphi Tr\{\overline{\Phi}_{i,j}^\varphi \cdot W_t\} \\ \sum_\varphi Tr\{\Phi_{i,j}^\varphi \cdot W_t\} & 1 & 0 \\ \sum_\varphi Tr\{\overline{\Phi}_{i,j}^\varphi \cdot W_t\} & 0 & 1 \end{bmatrix}$$
$$\succcurlyeq 0, (i,j) \in \Omega_l : \eta_{i,j,t}^l \quad (6b)$$
$$\begin{bmatrix} \varepsilon_v & Tr\{M_i^\varphi W_t\} - Tr\{M_i^\phi W_t\} \\ Tr\{M_i^\varphi W_t\} - Tr\{M_i^\phi W_t\} & \varepsilon_v \end{bmatrix}$$
$$\succcurlyeq 0, i \in \Omega_b, \varphi, \phi \in \{a,b,c\} : \beta_{i,\varphi,\phi t}^V \quad (6c)$$
where (1a) is converted to a semidefinite relaxation of (6a), and $\succcurlyeq 0$ denotes the corresponding positive semidefinite matrix. Similarly, (6b) and (6d) are the SDP relaxations of (5b) and (5i). Eq. (6c) enumerates voltage differences among the three phases corresponding to (5i).

*D. Compact Form of Sensitivity Matrices*

Here, we implement the sensitivity matrix analysis of system variables with respect to nodal voltages. Accordingly, we express the SDP-based three-phase unbalanced OPF explicitly with respect to the deviation of the nodal power injection.
$$\frac{\partial P_i^\varphi}{\partial X} = \frac{\partial Tr\{Y_i^\varphi W\}}{\partial X} = X^T\left(Y_i^\varphi + (Y_i^\varphi)^T\right) = JP_i^\varphi \quad (7a)$$
$$\frac{\partial Q_i^\varphi}{\partial X} = \frac{\partial Tr\{\overline{Y}_i^\varphi W\}}{\partial X} = X^T\left(\overline{Y}_i^\varphi + (\overline{Y}_i^\varphi)^T\right) = JQ_i^\varphi \quad (7b)$$
$$\frac{\partial |V_i^\varphi|^2}{\partial X} = \frac{\partial Tr\{M_i^\varphi W\}}{\partial X} = X^T\left(M_i^\varphi + (M_i^\varphi)^T\right) = JV_i^\varphi \quad (7c)$$
$$\frac{\partial P_{i,j}^\varphi}{\partial X} = \frac{\partial Tr\{\Phi_{i,j}^\varphi \cdot W\}}{\partial X} = X^T\left(\Phi_{i,j}^\varphi + (\Phi_{i,j}^\varphi)^T\right) = JP_{ij}^\varphi \quad (7d)$$
$$\frac{\partial Q_{i,j}^\varphi}{\partial X} = \frac{\partial Tr\{\overline{\Phi}_{i,j}^\varphi \cdot W\}}{\partial X} = X^T\left(\overline{\Phi}_{i,j}^\varphi + (\overline{\Phi}_{i,j}^\varphi)^T\right) = JQ_{ij}^\varphi \quad (7e)$$

We express the sensitivity matrices in a compact form. By expanding the row vector $JP_i^\varphi, JQ_i^\varphi \in \mathbb{R}^{1 \times 6N}$ into matrix $JP$, $JQ \in \mathbb{R}^{(3N \times 6N)}$, we have:
$$JP = [JP_1^a; JP_1^b; JP_1^c; \ldots; JP_i^\varphi; \ldots; JP_N^a; JP_N^b; JP_N^c] \quad (7f)$$
$$JQ = [JQ_1^a; JQ_1^b; JQ_1^c; \ldots; JQ_i^\varphi; \ldots; JQ_N^a; JQ_N^b; JQ_N^c] \quad (7g)$$

Herein, a power matrix $S = [P; Q]$ is introduced, where,
$$\frac{\partial S_t}{\partial X} = JS = [JP; JQ] \in \mathbb{R}^{(6N \times 6N)} \quad (7h)$$

Further, we use (7) in the following section to approximate the mapping of nodal uncertainties into system variables as a linear function.

### III. SDP-BASED DRCC-OPF MODEL FOR THREE-PHASE UNBALANCED DISTRIBUTION NETWORK

In the above SDP-based OPF model, nodes are characterized by their net active and reactive demands, defined as the difference between load demand and DER injections. As illustrated in the introduction, numerous literatures have pointed out that there are spatial correlations between uncertain variables in the distribution network [6], [18]. For the sake of generality, this paper analyzes the uncertainty of net load and defines the spatial correlation of uncertainties as the potential correlation among the three-phase net loads at different nodes. Since the renewable generation and load demand may deviate from their forecasts, the net demand deviation $\xi$ is stated as:

$$\Xi = \left\{ \mathbb{P}_\xi \left| \begin{array}{l} E(\xi) = \mu \\ E(\xi^T \xi) = \Gamma \end{array} \right. \right\}, \text{ where } \xi := \{\Delta d - \Delta w\} \quad (8a)$$

where $\Delta w$ and $\Delta d$ are nodal deviations of renewables and load demands. $\mu$ denotes the means and $\Gamma$ is the covariance matrix of uncertainties.

#### A. Modeling of spatially correlated uncertainties with information sharing mechanism

The Sklar's theorem states that the joint distribution any multivariate can be expressed as a set of univariate marginal distribution functions and a copula describing the dependence of variables [39]. Copula function $C$ is formulated by the CDF $F$ of multivariate, which is characterized by several marginal CDF $F_i$.

$$C[F_1(x_1), F_2(x_2), \cdots, F_n(x_n)] = F(x_1, x_2, \cdots, x_n) \quad (8b)$$

We develop an information sharing mechanism to describe the spatially correlated uncertainties via multivariate Gaussian copulas. Multivariate Gaussian copulas are referred to as elliptical copulas for representing the complicated relationships among variables [40]. The proposed mechanism makes use of shared information on the probability distribution of all uncertain variables using multivariate gaussian copulas. Accordingly,

$$C(u; \rho) = \Phi_N(\Phi^{-1}(u_1), \Phi^{-1}(u_2), \cdots, \Phi^{-1}(u_n)) \quad (8c)$$

where $u = (u_1, u_2, \cdots, u_n)^T$ are the uniform distributions between 0 and 1. $\Phi_N$ denotes the joint probability distribution function of $n$ standard normal distribution functions and $\Phi^{-1}$ indicates the inverse of the univariate standard normal CDF distribution. Covariance matrices can be constructed with the sensitivity coefficient $\rho$.

The Spearman correlation keeps invariant under CDFs and their inverse transformations. Accordingly, the rank correlation between variables remains the same when input variables are transformed from non-normal to normal distributions. Hence, with their CDFs, we develop a rank correlation function of any random multivariable:

$$\rho_{i,j}^s = \rho^s\left(F_i(x_i), F_j(x_j)\right) = \frac{cov(F_i(x_i), F_j(x_j))}{\sigma(F_i(x_i)) \cdot \sigma(F_j(x_j))} \quad (8d)$$

where $cov(a, b)$ represents the covariance of vector $a, b$ and $\sigma(a)$ is the standard deviation of vector $a$.

Given the uniform distribution limit of input data in multivariate Gaussian copula functions, we transform the uncertainty sources into their CDFs and then obtain the $(u_1, u_2, \cdots, u_n)$ variables, which follow uniform distributions located in the interval [0,1]. Based on this principle, the expected value, i.e., means $\mu$ for the spatial-correlated uncertain variables can be expressed by (8e). The inverse transformation method allows uncertain variables $R_m$ which follow arbitrary distributions to be expressed using a set of inverse CDFs together with the inverse function of copula as:

$$\mu = \mathbb{E}(R_m) = \mathbb{E}\left[F_m^{-1}(C^{-1}(\Phi_N))\right] \quad (8e)$$

Further, due to the fact that the covariance matrix $\Gamma$ can implicitly involve the standard deviations $\sigma$, we have:

$$\Gamma = (\sigma \cdot \sigma^T) \odot \rho \quad (8f)$$

In this case, $\rho := \{\rho_{i,j}\}$ is the matrices of Pearson correlation coefficients, and $\odot$ is an element-wise multiplication operator. For a joint distribution, $\rho_{i,j}$ is determined [40] as:

$$\rho_{i,j} = 2 \sin\left(\frac{\pi}{6} \cdot \rho_{i,j}^s\right) \quad (8g)$$

#### B. Nodal Deviation Matrix on Uncertainties

To mitigate forecast errors, balancing regulation capacity must be procured to match the electricity supply and demand. Given that the balancing regulation for total power mismatch caused by forecast errors must be compensated among distributed flexibility reserves, e.g., gas-fired turbines (GTs) and ESSs, a new set of auxiliary variables is introduced as $r_{i,\varphi} = \left(\beta_{i,t}^\varphi\right)^T \xi$ to quantify the relative system-wide forecast error that flexibility reserves at phase $\varphi$ of bus $i$ must compensate. $\beta_{i,t}^\varphi$ represents the column vector of balancing participation factors. The $k$th element $\beta_{i,\varphi}^k$ of $\beta_{i,t}^\varphi$ denotes the proportion of uncertainty source $\xi_k$ which will be balanced.

In accordance with the affine balance policy, the distribution of active power deviation across all nodes is modeled as a function of forecast errors and responded reserves as:

$$\Delta P = -A^\xi \cdot \xi + r = \left(-A^\xi + \beta^\xi\right) \cdot \xi \quad (9a)$$

where $\Delta P \in \mathbb{R}^{3N \times 1}$ denotes the power deviation in each phase of nodes. $A^\xi \in \mathbb{R}^{3N \times M}$, and $\beta^\xi \in \mathbb{R}^{3N \times M}$ are the incidence matrix for uncertainties and the matrix of balancing factors. M is the nodal value with uncertainties. For more clarity, we take the [3(i-1)+ $\varphi$]-th row of $\Delta P$ here.

$$\Delta P_{i,\varphi,t} = -A_{i,\varphi} \cdot \xi + r_{i,\varphi,t} = -A_{i,\varphi} \cdot \xi + \beta_{i,t}^\varphi \cdot \xi \quad (9b)$$

Coefficient matrices $A_{i,\varphi}$, $\beta_{i,t}^\varphi \in \mathbb{R}^{1 \times M}$ are constructed as:

$$A_{i,\varphi} := \left\{ a_i^\varphi \left| \begin{array}{l} a_i^\varphi = 1, \text{ if a load or DER is} \\ \text{located at bus } i \text{ on phase } \varphi \\ a_i^\varphi = 0, \text{ otherwise} \end{array} \right. \right\} \quad (9c)$$

$$\beta_{i,t}^\varphi = \left\{ \beta_{i,k,t}^\varphi \left| \begin{array}{l} \beta_{i,k,t}^\varphi \geq 0, \text{if a DG or ESS is located} \\ \text{at bus } i \text{ on phase } \phi \text{ for uncertainty } k \\ \beta_{i,k,t}^\varphi = 0, \text{ otherwise} \end{array} \right. \right\} \quad (9d)$$

#### C. System-wide Response Functions on Nodal Deviation

The system state variables cannot be derived directly due to the implicit nonlinear SDP-based OPF. Inspired by the affine policy [23], [24], this paper derives the response functions based on the first-order Taylor expansion to derive the expressions for the system-wide state variables, such as voltages and active/reactive power flows, affected by the uncertain nodal power injection [5]. However, different from the classical affine policy, the proposed response function analytically derives the mapping function of state variables relative to nodal injection power for any feasible operating point within the three-phase distribution network. We achieve this under the SDP-based formulation without approximating the nonlinear nature of the ACOPF. We offer following system-wide response functions:

**Proposition 1:** The response functions of voltages and active/reactive power flows to uncertainties are modeled as:

$$|\tilde{V}_{i,\varphi,t}|^2 = Tr\{\boldsymbol{M}_i^\varphi \boldsymbol{W}_t\} + \boldsymbol{XV}_{i,t}^\varphi \cdot \boldsymbol{\xi}_t \quad (10a)$$

$$\tilde{P}_{i,j,t}^\varphi = Tr\{\boldsymbol{\Phi}_{i,j}^\varphi \cdot \boldsymbol{W}_t\} + \boldsymbol{XP}_{ij,t}^\varphi \cdot \boldsymbol{\xi}_t \quad (10b)$$

$$\tilde{Q}_{i,j,t}^\varphi = Tr\{\bar{\boldsymbol{\Phi}}_{i,j}^\varphi \cdot \boldsymbol{W}_t\} + \boldsymbol{XQ}_{ij,t}^\varphi \cdot \boldsymbol{\xi}_t \quad (10c)$$

where the auxiliary matrix is listed as:

$$\boldsymbol{XV}_{i,t}^\varphi = \boldsymbol{JV}_{i,t}^\varphi \cdot \begin{bmatrix}\boldsymbol{JP}_t \\ \boldsymbol{JQ}_t\end{bmatrix}^{-1} \cdot \begin{bmatrix}-\boldsymbol{A}^\xi + \boldsymbol{\beta}_t^\xi \\ -\boldsymbol{\psi}\odot\boldsymbol{A}^\xi + \boldsymbol{\psi}\odot\boldsymbol{\beta}_t^\xi\end{bmatrix} \quad (10d)$$

$$\boldsymbol{XP}_{ij,t}^\varphi = \boldsymbol{JP}_{ij,t}^\varphi \cdot \begin{bmatrix}\boldsymbol{JP}_t \\ \boldsymbol{JQ}_t\end{bmatrix}^{-1} \cdot \begin{bmatrix}-\boldsymbol{A}^\xi + \boldsymbol{\beta}_t^\xi \\ -\boldsymbol{\psi}\odot\boldsymbol{A}^\xi + \boldsymbol{\psi}\odot\boldsymbol{\beta}_t^\xi\end{bmatrix} \quad (10e)$$

$$\boldsymbol{XQ}_{ij,t}^\varphi = \boldsymbol{JQ}_{ij,t}^\varphi \cdot \begin{bmatrix}\boldsymbol{JP}_t \\ \boldsymbol{JQ}_t\end{bmatrix}^{-1} \cdot \begin{bmatrix}-\boldsymbol{A}^\xi + \boldsymbol{\beta}_t^\xi \\ -\boldsymbol{\psi}\odot\boldsymbol{A}^\xi + \boldsymbol{\psi}\odot\boldsymbol{\beta}_t^\xi\end{bmatrix} \quad (10f)$$

The proof can be found in Appendix C. For all traces in (10a)-(10c) represent all feasible system state variables, which are formed in (2a)-(2c). The auxiliary matrices in (10d)-(10f) are actually the Jacobian matrix derived based on the SDP formulation. It does not depend on a given operating state, and the response function exists for any feasible operating state of the unbalanced distribution system.

### D. SDP-based DRCC-OPF Calculation in Unbalanced Three-Phase Power Distribution Network

The real-time variations of renewables and loads would cause unbalance in a three-phase distribution network. When balancing services are provided only by the upstream grid, the unbalanced operation could adversely affect the security of system operation without any local flexible resources in the three-phase distribution network. In practice, local flexible resources, such as distributed GTs and ESSs could provide operating reserves to absorb uncertainties within a certain range and, in turn, improve the operation security.

The dispatchable capability of a local flexible resource is determined by its allowable operating range and its predetermined reserve capacity. As shown in (9b), balancing regulation $\boldsymbol{r}$ compensated by flexible reserve resources is denoted as $\boldsymbol{r} = \boldsymbol{\beta} \cdot \boldsymbol{\xi}$. Assume the participation factors $\beta_{i,k,t}^\varphi$ must add up to over $\underline{\beta}$ ($0 \leq \underline{\beta} \leq 100\%$) in order to ensure that the network flexibility reserve can cope with power changes within at least $\beta$ ($0 \leq \beta \leq 100\%$) of the total uncertainties in (11a).

$$\sum_\varphi \sum_{i\in\Omega_g\cup\Omega_s} \beta_{i,k,t}^\varphi \geq \underline{\beta}: \lambda_{k,t}^R \quad (11a)$$

To ensure the activated balancing regulation of flexibility reserves could satisfy the energy dispatch and flow limits even in the worst case, the following DR-based chance constraints are modeled into the original three-phase unbalanced OPF model with a probability of $1-\epsilon$, where $\epsilon$ represents the confidence level (probability) set by DSO.

$$inf_{P_\xi\in\Xi}\mathbb{P}_\xi\{R_{i,t}^{up} \geq r_{i,t}\} \geq 1 - \epsilon_R \quad (11b)$$

$$inf_{P_\xi\in\Xi}\mathbb{P}_\xi\{R_{i,t}^{dn} \geq -r_{i,t}\} \geq 1 - \epsilon_R \quad (11c)$$

$$inf_{P_\xi\in\Xi}\mathbb{P}_\xi\{|\tilde{V}_{i,t}^\varphi|^2 \leq (V_i^{max})^2\} \geq 1 - \epsilon_v \quad (11d)$$

$$inf_{P_\xi\in\Xi}\mathbb{P}_\xi\{|\tilde{V}_{i,t}^\varphi|^2 \geq (V_i^{min})^2\} \geq 1 - \epsilon_v \quad (11e)$$

$$inf_{P_\xi\in\Xi}\mathbb{P}_\xi\{|\sum_\varphi \tilde{P}_{i,j,t}^\varphi| \leq t_{ij,t}^p\} \geq 1 - \epsilon_f \quad (11f)$$

$$inf_{P_\xi\in\Xi}\mathbb{P}_\xi\{|\sum_\varphi \tilde{Q}_{i,j,t}^\varphi| \leq t_{ij,t}^q\} \geq 1 - \epsilon_f \quad (11g)$$

$$(t_{ij,t}^p)^2 + (t_{ij,t}^q)^2 \leq (S_{ij}^{max})^2, \forall(i,j)\in\Omega_l: \bar{\alpha}_{l,t}^{line} \quad (11h)$$

where $\mathbb{P}_\xi\{\cdot\}$ denotes a probability distribution function of the uncertainties $\xi$ from the set of possible distributions $\Xi$.

**Remark 1:** Although chance constraints (11b)-(11g) are nonconvex and intractable, we can make a second-order cone approximation using the proposed system-wide response functions (10a)-(10b). Notice that the reformulations are sufficient conditions for the original model according to the Chebyshev inequality [5], [30].

**Proposition 2:** For any random variables, follow the normal distribution $\boldsymbol{X} \sim Norm(\boldsymbol{\mu}, \boldsymbol{\Gamma})$, the inequality $\mathbb{P}(X \leq x^{max}) \geq \eta$ holds if and only if $x^{max} \geq \mu + z_\eta\Gamma^{1/2}$, where $z_\eta := \Phi^{-1}(1-\eta)$ is the $(1-\eta)$-quantile of a standard normal distribution [41].

**Remark 2:** Proposition 2 relies on the assumption that all transformed variables follow a normal distribution; however, uncertainty variables might not be normally distributed in their original form. Instead, we use the information-sharing mechanism to convert the original distribution to derive the expectation and covariance matrix of spatially-coupled uncertainty variables, as shown in (8e) and (8f).

Herein, we recast the intractable (11b)-(11g) as the following reformulation based on the above sensitivity analysis.

- **Reformulation for the DRCC of reserve dispatch**

$$R_{i,t}^{up} \geq \sum_\varphi (\boldsymbol{\beta}_{i,t}^\varphi)^T \cdot \boldsymbol{\mu}_t + z_R \left\|\boldsymbol{\Gamma}_t^{\frac{1}{2}}\boldsymbol{\beta}_{i,t}^\varphi\right\|_2, i\in\Omega_S\cup\Omega_g: \bar{\alpha}_{i,t}^R \quad (12a)$$

$$R_{i,t}^{dn} \geq \sum_\varphi z_R \left\|\boldsymbol{\Gamma}_t^{\frac{1}{2}}\boldsymbol{\beta}_{i,t}^\varphi\right\|_2 - (\boldsymbol{\beta}_{i,t}^\varphi)^T \cdot \boldsymbol{\mu}_t, i\in\Omega_S\cup\Omega_g: \underline{\alpha}_{i,t}^R \quad (12b)$$

where $\boldsymbol{\Gamma}_t^{1/2}$ is a matrix satisfying: $(\boldsymbol{\Gamma}_t^{1/2})^T \boldsymbol{\Gamma}_t^{1/2} = \boldsymbol{\Gamma}$, which shows the positive semi-definiteness of $\boldsymbol{\Gamma}$. The uncertainty margin factor $z_R = \sqrt{\frac{1-\epsilon_R}{\epsilon_R}}$ is constant, which is determined by the risk tolerance.

- **Reformulation for the DRCC of voltage security**

$$(V_i^{max})^2 \geq E\left(|\tilde{V}_{i,\varphi,t}|^2\right) + z_v \cdot Stdev\left(|\tilde{V}_{i,\varphi,t}|^2\right): \bar{\alpha}_{i,\varphi,t}^V \quad (12c)$$

$$(V_i^{min})^2 \leq E\left(|\tilde{V}_{i,\varphi,t}|^2\right) - z_v \cdot Stdev\left(|\tilde{V}_{i,\varphi,t}|^2\right): \underline{\alpha}_{i,\varphi,t}^V \quad (12d)$$

$$E\left(|\tilde{V}_{i,\varphi,t}|^2\right) = Tr\{\boldsymbol{M}_i^\varphi \boldsymbol{W}_t\} + \boldsymbol{XV}_{i,t}^\varphi \cdot \boldsymbol{\mu}_t \quad (12e)$$

$$Stdev\left(|V_{i,\varphi,t}(\xi)|^2\right) = \|\boldsymbol{XV}_{i,t}^\varphi \cdot \boldsymbol{\Gamma}_t^{1/2}\|_2 \quad (12f)$$

where $z_v = \Phi^{(-1)}(1-\epsilon_v)$ represents the uncertainty margin factor for voltage security, $E\left(|\tilde{V}_{i,\varphi,t}|^2\right)$ and $Stdev\left(|\tilde{V}_{i,\varphi,t}|^2\right)$ are the expectation and standard deviation of the nodal voltage magnitude squared under uncertainty.

- **Reformulation for the DRCC of power flow**

$$t_{i,j,t}^p + E\left(\sum_\varphi \tilde{P}_{i,j,t}^\varphi\right) \geq z_l \cdot Stdev\left[\sum_\varphi \tilde{P}_{i,j,t}^\varphi\right]: \bar{\alpha}_{l,t}^p \quad (12g)$$

$$t_{i,j,t}^p - E\left(\sum_\varphi \tilde{P}_{i,j,t}^\varphi\right) \geq z_l \cdot Stdev\left[\sum_\varphi \tilde{P}_{i,j,t}^\varphi\right]: \underline{\alpha}_{l,t}^p \quad (12h)$$

$$t_{i,j,t}^q + E\left(\sum_\varphi \tilde{Q}_{i,j,t}^\varphi\right) \geq z_l \cdot Stdev\left[\sum_\varphi \tilde{Q}_{i,j,t}^\varphi\right]: \bar{\alpha}_{l,t}^q \quad (12i)$$

$$t_{i,j,t}^q - E\left(\sum_\varphi \tilde{Q}_{i,j,t}^\varphi\right) \geq z_l \cdot Stdev\left[\sum_\varphi \tilde{Q}_{i,j,t}^\varphi\right]: \underline{\alpha}_{l,t}^q \quad (12j)$$

$$E\left(\sum_\varphi \tilde{P}_{i,j,t}^\varphi\right) = \sum_\varphi P_{i,j,t}^\varphi + \sum_\varphi \boldsymbol{XP}_{ij,t}^\varphi \cdot \boldsymbol{\mu}_t \quad (12k)$$

$$E\left(\sum_\varphi \tilde{Q}_{i,j,t}^\varphi\right) = \sum_\varphi Q_{i,j,t}^\varphi + \sum_\varphi \boldsymbol{XQ}_{ij,t}^\varphi \cdot \boldsymbol{\mu}_t \quad (12l)$$

$$Stdev\left[\sum_\varphi \tilde{P}_{i,j,t}^\varphi\right] = \left\|\sum_\varphi \boldsymbol{XP}_{ij,t}^\varphi \cdot \boldsymbol{\Gamma}_t^{1/2}\right\|_2 \quad (12m)$$

$$Stdev\left[\sum_\varphi \tilde{Q}_{i,j,t}^\varphi\right] = \left\|\sum_\varphi \boldsymbol{XQ}_{ij,t}^\varphi \cdot \boldsymbol{\Gamma}_t^{1/2}\right\|_2 \quad (12n)$$

where $z_l = \Phi^{(-1)}(1-\epsilon_l)$ represents the uncertainty margin factor for power flow. $E(\sum_\varphi \tilde{P}_{i,j,t}^\varphi)$, $E(\sum_\varphi \tilde{Q}_{i,j,t}^\varphi)$ are the expectation for the sum of active/reactive power flows on all phases. Meanwhile, $Stdev[\sum_\varphi \tilde{P}_{i,j,t}^\varphi]$ and $Stdev[\sum_\varphi \tilde{Q}_{i,j,t}^\varphi]$ are the standard deviations for real and reactive power flows, respectively.

Substituting the above equivalent constraints (12) for chance constraints (11b)-(11g), the original DRCC three-phase OPF for the unbalanced power distribution model will be reformulated into an SDP-based OPF model as:

$$\min \sum_{i,t} C_{i,P} + \sum_i \left(C_{i,R}^{\text{up}}(R_i^{\text{up}}) + C_{i,R}^{\text{dn}}(R_i^{\text{dn}})\right)$$

s.t.      Generation and ramping limits (4a);
ESSs charging/discharging limits (4b);
Three-phase power flow limits: (2a)-(2e), (5a-5h);
SDP relaxation: (6);
Reserve activation constraints: (11a)
Reformulated DR chance constraints: (12), (11h)

Fig.1 demonstrates the flowchart of the proposed risk-aware flexible resource utilization model. The uncertainty analysis is made at the first stage, which integrates information-sharing methods to model the spatially-correlated uncertainties. In the second stage, DSO will perform the flexibility market-clearing framework to derive the risk-aware flexibility prices. The market-clearing framework involves three steps: First, we establish an SDP-based ACOPF model for an unbalanced three-phase distribution network. Then, a DRCC model is formulated to assess the impact of spatially-correlated uncertainty on the network operation. In this way, a flexibility market is developed to quantify the value of local flexibility reserves. Finally, based on the duality theory, we derive the risk-aware flexibility prices from the SDP-based DRCC-ACOPF model.

## IV. ANALYTIC FRAMEWORK FOR THE CLEARED RISK-AWARE FLEXIBILITY PRICES

The DLMP for energy can be derived from the dual multipliers $\lambda_{i,\varphi,t}^P$ and $\lambda_{i,\varphi,t}^Q$ in (5e)-(5f) using the DRCC model for the solution of the unbalanced three-phase ACOPF,. Similar results can be obtained for the risk-aware flexibility prices provided by up/down reserves $\bar{\alpha}_{i,t}^R$ and $\underline{\alpha}_{i,t}^R$, which are derived from the dual multipliers in (12a)-(12b). As both upward and downward reserves are functions of the balancing factor $\beta_{i,t}$, their marginal prices are coupled. The DSO's cleared risk-aware flexibility pricing scheme is performed on both sides of the flexibility market by charging the uncertainties and rewarding flexible resources simultaneously. In this section, we first decompose flexibility prices for rewarding flexible resources to demonstrate their physical meaning. Then, we derive the analytical decomposition of cleared flexibility pricing prices for charging the uncertainties.

### A. Composition of DSO's Cleared Risk-Aware Flexibility Prices for Rewarding the Local Flexible Resources

Based on shadow prices presented in the duality theory [5], the cleared risk-aware flexibility price is a compensation for the marginal cost of uncertainty. Given the Lagrange function of the DRCC model of the unbalanced three-phase ACOPF, the rewarding prices for local flexible resources are obtained by taking the first order partial derivatives of $\beta_{i,k,t}^\varphi$ to zero.

***Proposition 3***: When the mean of uncertain deviation is zero, i.e., $E(\boldsymbol{\mu})=0$, we can decompose the flexibility pricing into four parts: 1) energy flexibility cost; 2) voltage flexibility cost; 3) active power flow flexibility cost; 4) reactive power flow flexibility cost. The formulation is derived as:

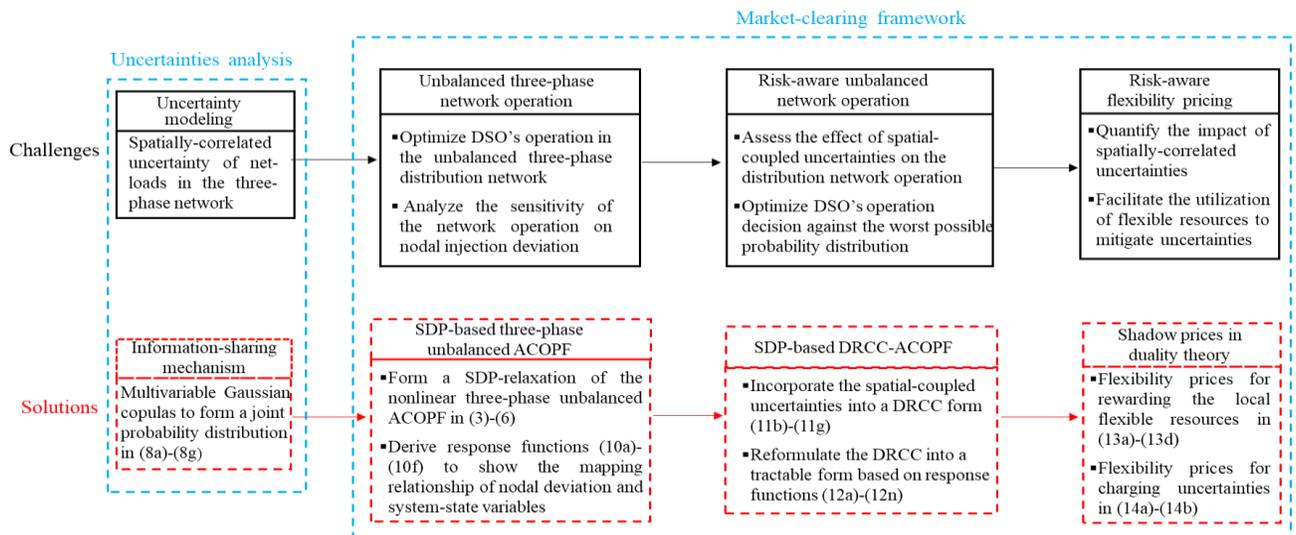

Fig. 1. Flowchart of the proposed risk-aware flexible resource utilization model.

$$\left(\bar{\alpha}_{i,t}^R + \underline{\alpha}_{i,t}^R\right) = \frac{\left\|r_t^{1/2} \cdot \beta_{i,t}^\varphi\right\|_2}{z_R \cdot (e_k)^T \cdot r_t \cdot \beta_{i,t}^\varphi} *$$

$$\left\{ \begin{array}{l} \lambda_{t,k}^R - \sum_{i,\varphi}\left(\bar{\alpha}_{i,\varphi,t}^V + \underline{\alpha}_{i,\varphi,t}^V\right) \cdot Z_{V_{i,t}^\varphi} \\ -\sum_{ij \in L}\left(\left(\bar{\alpha}_{ij,t}^p + \underline{\alpha}_{ij,t}^p\right) \cdot Z_{P_{ij,t}} - \left(\bar{\alpha}_{ij,t}^q + \underline{\alpha}_{ij,t}^q\right) \cdot Z_{Q_{ij,t}}\right) \end{array} \right\} \quad (13a)$$

where $\lambda_{t,k}^R$, $\bar{\alpha}_{i,\varphi,t}^V$, $\underline{\alpha}_{i,\varphi,t}^V$, $\bar{\alpha}_{i,j,t}^p$, $\underline{\alpha}_{i,j,t}^p$, $\bar{\alpha}_{i,j,t}^q$ and $\underline{\alpha}_{i,j,t}^q$ are the Lagrangian multipliers in (11a), (12c)-(12d), (12g)-(12j). $Z_{V_{i,t}^\varphi}$, $Z_{P_{i,t}^\varphi}$ and $Z_{Q_{i,t}^\varphi}$ are auxiliary variables, defined as:

$$Z_{V_{i,t}^\varphi} = \frac{z_v \cdot XV_{i,t}^\varphi \cdot \Gamma_t \cdot e_k \cdot JV_{i,t}^\varphi \left[\begin{array}{c} JP_t \\ JQ_t \end{array}\right]^{-1} \cdot \left[\begin{array}{c} e_i^\varphi \\ \psi \odot e_i^\varphi \end{array}\right]}{\left\|XV_{i,t}^\varphi \cdot \Gamma_t^{1/2}\right\|_2} \quad (13b)$$

$$Z_{P_{ij,t}} = \frac{z_l \cdot \sum_\varphi XP_{ij,t}^\varphi \cdot \Gamma_t \cdot e_k \cdot JP_{ij,t}^\varphi \left[\begin{array}{c} JP_t \\ JQ_t \end{array}\right]^{-1} \left[\begin{array}{c} e_i^\varphi \\ \psi \odot e_i^\varphi \end{array}\right]}{\left\|\sum_\varphi XP_{ij,t}^\varphi \cdot \Gamma_t^{1/2}\right\|_2} \quad (13c)$$

$$Z_{Q_{i,t}} = \frac{z_l \cdot \sum_\varphi XQ_{ij,t}^\varphi \cdot \Gamma_t \cdot e_k \cdot JQ_{ij,t}^\varphi \left[\begin{array}{c} JP_t \\ JQ_t \end{array}\right]^{-1} \left[\begin{array}{c} e_i^\varphi \\ \psi \odot e_i^\varphi \end{array}\right]}{\left\|\sum_\varphi XQ_{ij,t}^\varphi \cdot \Gamma_t^{1/2}\right\|_2} \quad (13d)$$

The first term $\lambda_{t,k}^R$ in (13a) denotes the system-wide flexibility reserve cost for the uncertainty source k, and $\left(\bar{\alpha}_{i,\varphi,t}^V + \underline{\alpha}_{i,\varphi,t}^V\right) \cdot Z_{V_{i,t}^\varphi}$ is the voltage-dependent part of reserve cost. Further, $\left(\bar{\alpha}_{i,j,t}^p + \underline{\alpha}_{i,j,t}^p\right) \cdot Z_{P_{i,t}}$ and $\left(\bar{\alpha}_{i,j,t}^q + \underline{\alpha}_{i,j,t}^q\right) \cdot Z_{Q_{i,t}}$ is the power flow part of reserve cost. The proof of *Proposition 3* is provided in the Appendix D.

*B. Composition of DSO's Cleared Risk-Aware Flexibility Prices for Charging Uncertainties*

In this section, we derive the composition of the DSO's cleared risk-aware flexibility prices for charging uncertainties. The shadow price in duality theory is further employed to quantify the marginal cost on the network operation brought by uncertainties. The uncertain charging scheme incentivizes flexible resources to mitigate uncertain fluctuations locally.

***Proposition 4***: The cleared flexibility prices for charging the uncertainties is derived as partial derivatives of the Lagrangian function. The charging scheme is represented by the means and the standard deviations of net demand forecast error. It should be noted that $\lambda_{\mu,k,t}$ and $\lambda_{\sigma,k,t}$ can be decomposed into four parts: 1) energy; 2) volt/var; 3) active power flow; 4) reactive power flow:

$$\lambda_{\mu,k,t} := \frac{\partial \mathcal{L}}{\partial \mu_{k,t}} = \sum_{i,\varphi}\left(\bar{\alpha}_{i,t}^R - \underline{\alpha}_{i,t}^R\right) \cdot \beta_{i,k,t}^\varphi$$
$$+ \sum_{i,\varphi}\left(\bar{\alpha}_{i,\varphi,t}^V - \underline{\alpha}_{i,\varphi,t}^V\right) \cdot XV_{i,k,t}^\varphi$$
$$+ \sum_{l,\varphi}\left(\bar{\alpha}_{l,t}^p - \underline{\alpha}_{l,t}^p\right) \cdot XP_{l,k,t}^\varphi + \sum_{l,\varphi}\left(\bar{\alpha}_{l,t}^q - \underline{\alpha}_{l,t}^q\right) \cdot XQ_{l,k,t}^\varphi \quad (14a)$$

$$\lambda_{\sigma,k,t} := \frac{\partial \mathcal{L}}{\partial \sigma_k} = \sum_{i,\phi}\left(\bar{\alpha}_{i,t}^R + \underline{\alpha}_{i,t}^R\right) \cdot \frac{z_R \cdot \sum_{k*} \beta_{i,k,t}^\varphi \cdot \beta_{i,k*,t}^\varphi \cdot \sigma_{k*}}{\left\|r_t^{1/2} \cdot \beta_{i,t}^\varphi\right\|_2}$$
$$+ \sum_{i,\varphi}\left(\bar{\alpha}_{i,\varphi,t}^V + \underline{\alpha}_{i,\varphi,t}^V\right) \cdot \frac{z_v \cdot \sum_{k*} XV_{i,k,t}^\varphi \cdot XV_{i,k*,t}^\varphi \cdot \sigma_{k*}}{\left\|XV_{i,t}^\varphi \cdot \Gamma_t^{1/2}\right\|_2}$$
$$+ \sum_{l,k*}\left(\bar{\alpha}_{l,t}^p + \underline{\alpha}_{l,t}^p\right) \cdot \frac{z_l \cdot \sum_\varphi XP_{l,k,t}^\varphi \cdot \sum_\varphi XP_{l,k*,t}^\varphi \cdot \sigma_{k*}}{\left\|\sum_\varphi XP_{l,t}^\varphi \cdot \Gamma_t^{1/2}\right\|_2}$$
$$+ \sum_{l,k*}\left(\bar{\alpha}_{l,t}^q + \underline{\alpha}_{l,t}^q\right) \cdot \frac{z_l \cdot \sum_\varphi XQ_{l,k,t}^\varphi \cdot \sum_\varphi XQ_{l,k*,t}^\varphi \cdot \sigma_{k*}}{\left\|\sum_\varphi XQ_{l,t}^\varphi \cdot \Gamma_t^{1/2}\right\|_2} \quad (14b)$$

where $\lambda_{\mu,k,t}$ and $\lambda_{\sigma,k,t}$ are uncertain prices for $\mu_{k,t}$ and $\sigma_{k,t}$, respectively. The first term is related with the marginal energy cost for mitigating uncertainties, the second to the last term denote the voltage reserve cost, and active and reactive branch power flow reserve costs.

***Remark 3:*** The uncertainty charging prices for standard deviations show a strong correction with the system-wide uncertainty level, i.e., the sum of all standard deviations $\sigma_{k*}$. Given that the information-sharing mechanism can reduce the system-wide forecast error, the total cost saving is included in the price signal. The cleared flexibility prices for both sides of the DSO's flexibility market based on information sharing can effectively promote information transparency, while allowing the flexible sources to mitigate uncertainty costs by improving the quality of forecast and investing in local flexible resources.

*C. Analysis of Money Flow and Profit Sufficiency Guarantee*

According to the proposed market-clearing framework, uncertainty sources should compensate flexible resources for the provision of reserve costs. Based on the definition of a flexibility pricing scheme, the flexibility provider's revenue $C_{i,t}^R$ is calculated as:

$$C_{i,t}^R := \bar{\alpha}_{i,t}^R \cdot R_{i,t}^{\text{up}} + \underline{\alpha}_{i,t}^R \cdot R_{i,t}^{\text{dn}} \quad (15a)$$

Similarly, the uncertainty payment to the flexible resource provider is calculated as:

$$E_{k,t}^\xi := \lambda_{\mu,k,t} \cdot \mu_k + \lambda_{\sigma,k,t} \cdot \sigma_{k,t} \quad (15b)$$

Based on (15a)-(15b), we offer the following proposition.

***Proposition 5 (Profit sufficiency)***: The uncertainty payment $\sum_k E_{k,t}^\xi$ would cover the total flexibility cost $\sum_i C_{i,t}^R$. Meanwhile, the profit sufficiency reflects the compensation for the network operation margin. This proof is provided in Appendix E.

## V. CASE STUDY

The proposed DSO's risk-aware flexibility market clearing scheme is verified by a modified IEEE 34-bus test system. The proposed model is implemented in the YALMIP toolbox with Mosek as the SDP solver [42]. The numerical computation is carried out on a personal computer with 12th Gen Intel (R) Core (TM) i7-12700 (4.70 GHz).

*A. Test Case Description*

The modified IEEE 34-bus three-phase distribution network system, shown in Fig. 2, is used to validate the proposed risk-aware local flexibility and uncertainty pricing scheme. The system includes four WTs, three GTs, and three distributed ESSs. Some critical DER parameters are provided in Tables I and II. All WTs located in Nodes 8, 21, 26, and 32 have a capacity of 0.15 MW.

TABLE I
ESS parameters

| Node | $SOC_i^{max}$ | $ch_i^{max}$ | $dis_i^{max}$ | $\eta_s$ | $b_1$ | $b_0$ |
|------|---------------|--------------|---------------|----------|-------|-------|
| 8    | 0.20          | 0.15         | 0.15          |          | 0.11  |       |
| 11   | 0.52          | 0.42         | 0.42          | 0.9      | 0.10  | 0     |
| 26   | 0.20          | 0.15         | 0.15          |          | 0.19  |       |

* $SOC_i^{max}$, $ch_i^{max}$ / $dis_i^{max}$, $\eta_s$: max capacity (MWh), power charge/discharge (MW) and efficiency
* Operation cost: $b_1*|ch - dis|+b_0$ ($)

TABLE II
GT parameters

| Node | $g_i^{min}/g_i^{max}$ | $RU_i$ | $RD_i$ | $PF_i^{min}/PF_i^{max}$ | $a_2$ | $a_1$ |
|---|---|---|---|---|---|---|
| 6 | 0/0.84 | 0.6 | 0.6 |  | $8*10^{-3}$ | 30 |
| 9 | 0/0.72 | 0.4 | 0.4 | 0.1/0.9 | $1.1*10^{-2}$ | 35 |
| 19 | 0/0.96 | 0.8 | 0.8 |  | $1.2*10^{-2}$ | 40 |

\* $g_i^{max}$, $RU_i/RD_i$, $PF_i^{min}/PF_i^{max}$: max power output (MW), up/down ramping rate (MW/h) and minimum/maximum power factor
\* Operation cost: $a_1*g + a_2*g^2$ ($)

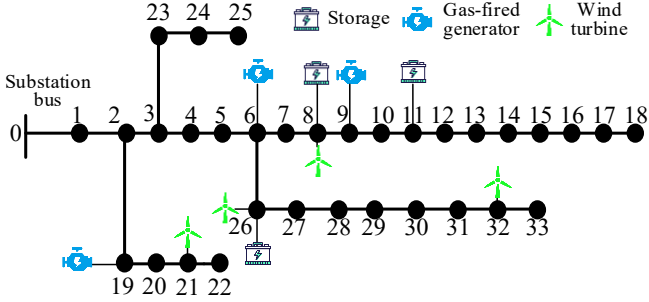

Fig. 2. Modified IEEE 34-bus distribution system.

The wind and load forecast data are provided in [43], [44] and the time of use tariff for electricity $\lambda^{TOU}$ is provided in [43]. The prices for the reactive power obtained from the ancillary services market and flexibility resources are set as $0.2*\lambda^{TOU}$. The reserve price provided by flexibility resources depends on their bids [45]. The security three-phase voltage ranges from 0.9 p.u. to 1.1 p.u, while the maximum unbalanced voltage deviation index (VDI) $\varepsilon_v$ is set as 10%, which can be adjusted by distribution network operator.

*B. Market Clearing Result*

Herein, we analyze the merits of the proposed risk-aware pricing scheme by quantifying the effect of incentive on operation costs and the improvement on the operation performance as stated in Tables III and IV, respectively. Three schemes are designed to illustrate the effectiveness of the proposed method, where Schemes I and II are used as benchmarks.

Scheme I (Stochastic solution one without a DSO's flexibility market): All uncertainties will be balanced by the upstream grid without utilizing local flexible resources. The day-ahead dispatch is executed by a multi-scenario stochastic method [46].

Scheme II (Chance-constrained linear robust solution without information sharing): A chance-constrained linear robust method is used to form a risk-aware price as a benchmark [30]. The information sharing for spatially correlated uncertainty is not considered in this case. The DSO develops the risk-aware pricing scheme based on estimated system-wide uncertainties, i.e., forecast of total net demand fluctuation.

Scheme III (DRCC solution proposed in this paper): The proposed DRCC risk-aware pricing scheme coordinates the spatially correlated uncertainty information. The DSO purchases the local flexible reserve using the shared uncertainty information.

Table III compares the operation cost of different schemes in the worst after-the-fact situation. The worst after-the-fact situation implies the most dramatic fluctuations in uncertainty, where the real-time uncertainty emerges with minimum renewable output and maximum load demand. In Table III, the first two cost terms include the day-ahead electricity and reserve purchase costs and the real-time balancing cost for the worst-case deviation.

TABLE III
Effect of incentive on operation costs ($)

| Scheme | $C^{sub,P}$ ($) | $C^{sub,Q}$ ($) | $C^{DER}$ ($) | $C^{UP} + C^{DN}$ ($) | total ($) |
|---|---|---|---|---|---|
| I | 396.4 | 82.1 | 298.8 | - | 777.3 |
| II | 278.8 | 65.4 | 232.6 | 173.3 | 750.1 |
| III | 257.6 | 64.6 | 244.1 | 78.2 | 644.5 |

Note that $C^{sub,P}$, $C^{sub,Q}$ are purchasing costs for active and reactive energy from the external market. $C^{DER}$ denotes the total dispatch cost for DERs while $C^{UP} + C^{DN}$ refers to the purchasing cost for up/down local flexibility reserves.

The proposed method results in the lowest total operation cost of $644.5, which is reduced by 17.1% as compared with Scheme I because active and reactive energy are purchased from the market. Here, GTs will have more capacity available for participation in the day-ahead energy market with a higher generation cost as there is no need for the local flexibility reserve in Scheme I. Similarly, GTs would participate less in the day-ahead energy market as additional flexible reserves are utilized, which in turn leads to a lower generation cost.

Table IV shows the improvement offered by the proposed scheme on the network operation in the worst after-the-fact situation. Herein, $\hat{S}_l^{max}$ refers to the maximum line loading (i.e., maximum line power flow on the day-ahead time scale in % of line capacity). The second term denotes the nodal voltage magnitude range in the worst case. $VDI_i^{max}$ is the largest unbalanced deviation index of the three-phase nodal voltage.

Also, certain operation constraints are not satisfied in Scheme I, as it fails to invoke flexible resources in the unbalanced three-phase distribution network. In the worst after-the-fact situation, uncertain deviations cannot be balanced by the upstream grid alone without violating the distribution network constraints. Conversely, both Scheme II and the proposed Scheme III have a good performance in mitigating line congestion, voltage fluctuation, and three-phase unbalanced level. However, the proposed Scheme III improves the distribution network operation performance to be almost the same level as that of Scheme II at a cheaper cost.

TABLE IV
Comparison of operation performances

| Scheme | $\hat{S}_l^{max}$ | $|\tilde{V}_{i,\varphi}^{min}| \sim |\tilde{V}_{i,\varphi}^{max}|$ | $VDI_i^{max}$ |
|---|---|---|---|
| I | 114.6% | 0.839~1.026 | 13.42% |
| II | 94.8% | 0.956~1.012 | 5.53% |
| III | 97.8% | 0.944~1.024 | 7.02% |

In Fig. 3, a cost-revenue analysis is conducted with and without sharing spatially correlated uncertainties in order to evaluate the economic implications of a risk-aware three-phase pricing scheme for flexible resources. The comparison of these two schemes shows that distributed WT revenues and cost saving for loads are raised from $5.56 and $-329.68 to $57.64 and $-289.28, respectively. The total net profit for all participants will increase despite an approximately 6.58%-13.09% reduction in profit for distributed GTs and ESSs. The reason is that sharing spatially correlated uncertainties can reduce the total flexibility reserve requirement. Concretely, GT and ESS revenues are somewhat lower than those without sharing information of spatially correlated uncertainties, while the proposed solution reduces the risk payment by distributed WTs and loads. On the other hand, ESS revenues are primarily derived from offering reserve credits rather than arbitraging on electricity sales.

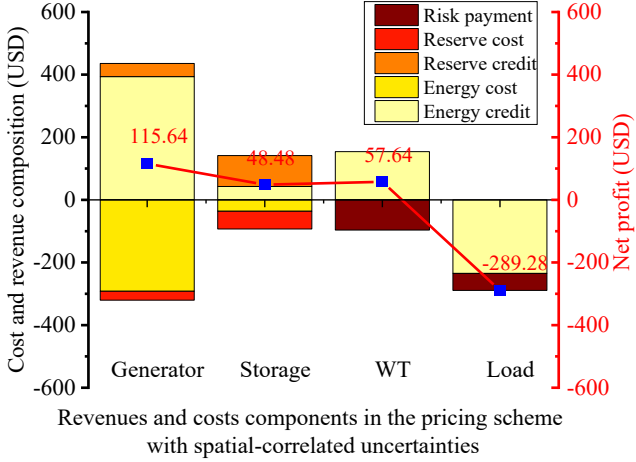

Fig. 3. (a) Comparison of revenue and cost components with and without the information sharing on spatially correlated uncertainties

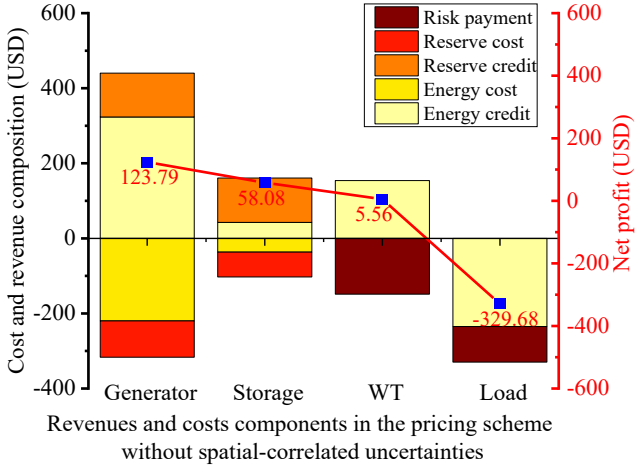

Fig. 3. (b) Comparison of revenue and cost components with and without the information sharing on spatially correlated uncertainties

Basically, the reserve is mainly provided by ESSs for their cheap flexibility. Meanwhile, the provision of reserve is less profitable than electricity sales for GTs, which is validated by Fig. 4. The distributed GT dispatch is more suited for arbitrage, where the results vary with fluctuations in external electricity prices. GTs will be dispatched, as market price is increased, for lowering the power purchase cost as demonstrated in 13th-22th timeslots. ESSs will be charged (i.e., negative values) as renewable energy production is increased or the price of electricity is decreased; otherwise, ESSs are discharged (i.e., positive value). However, the ESS dispatch accounts for only a small portion of the total supplied electricity. Accordingly, ESS should possess a sufficient reserve capacity for providing flexible reserves.

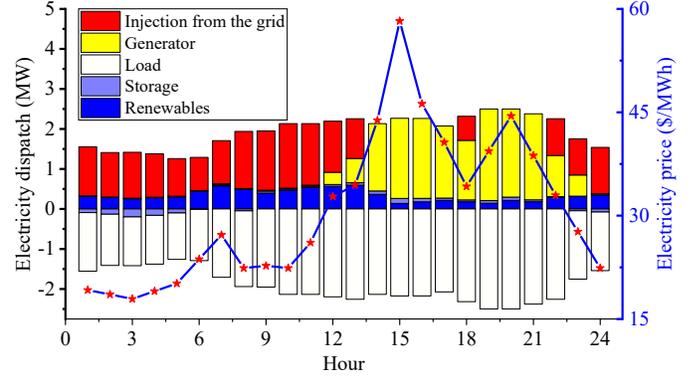

Fig. 4. Electricity dispatch and external wholesale market price

In our study, we have carefully analyzed the accuracy of the SDP relaxation method. For the SDP-based model, the solution is exact when the rank-one condition of the matrix $W_t$ defined in Eq. (1a) is satisfied [47]. However, the rank condition is usually not satisfied due to the relaxation. Since the matrix's rank, which is the number of the nonzero singular values, provides the information about the accuracy of the solution. Fig. 5 (a) shows the eigenvalues of the matrix $W_t$ on a logarithm scale with different thresholds of the voltage stability, while Fig. 5 (b) denotes the ratios between the largest and second-largest Eigenvalues of the matrix $W_t$ under different voltage deviation index. The red line in the box plot of Fig. 5 (a) indicates the median of the samples, the dashed lines represent the 5%-95% interval, and the box represents the 25%-75% interval.

The results show that the distribution of eigenvalues is not uniform, i.e., only a few top eigenvalues are large and the rest are of small magnitudes. The results show that the smaller the voltage deviation index, the larger the overall eigenvalues. Fig. 5 (b) shows that there is one large singular value, and the other singular values are so small that they can be ignored compared to the largest singular value. This indicates that the rank of the matrix W can be approximately considered to be 1 [48]. The smaller the three-phase voltage deviation margin is, the solution will be more exact. Case studies show that the proposed SDP relaxation method can provide accurate solutions for the nonconvex three-phase ACOPF problem under practical operating conditions.

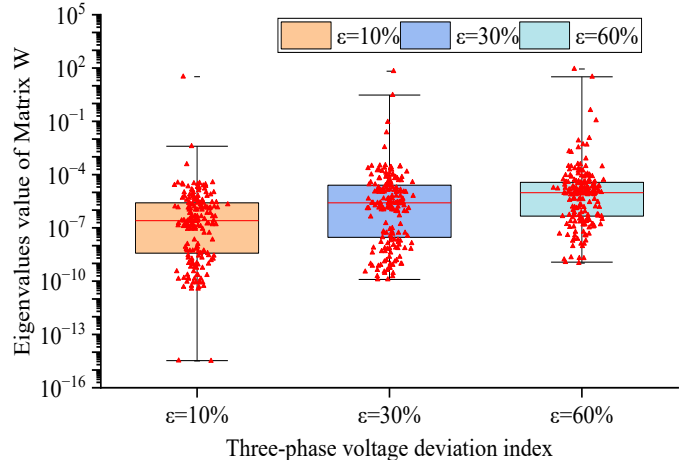

Fig. 5. (a) Eigenvalues of matrix under diverse voltage deviation index

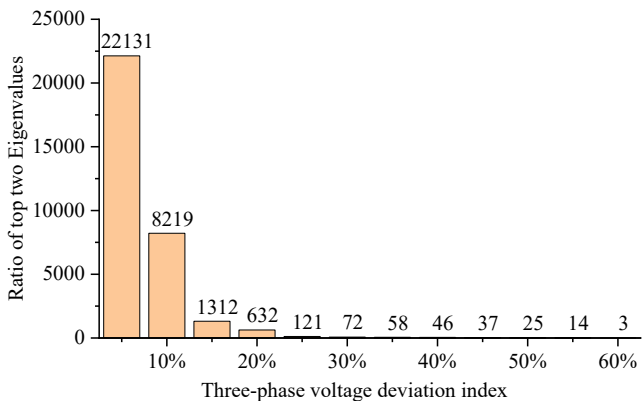

Fig. 5. (b) Ratio of the Top 2 eigenvalues under diverse voltage dexiation index

Fig. 6 shows the day-ahead market clearing results for local up/down reserves, where three conclusions can be derived. First, the ESS reserve price tends to be lower than that of GT. This is why ESSs are the main reserve providers, even with a relatively smaller installed capacity. Another implication is that the ESS price for the upward reserve (averaged at 4.12 $/MWh) tends to be higher than that of the downward reserve, with a mean value of 2.82 $/MWh. This is because the provision of upward reserve often leads to higher energy costs. Third, the cleared reserve price for GT varies dramatically. The GT reserve price is typically determined by two factors: market price and total system-wide flexibility reserve requirement. Especially, when the system-wide requirement for down reserve exceeds ESS capacities, GTs would increase their power outputs to offer more downward reserves. In Fig. 4, down reserve prices for GTs are much higher than those of upward when the GTs are less dispatched at 1st to 12th and 23rd to 24th timeslots. In the gray area, the upward reserve price is higher than that of the downward reserve. This is also owing to a higher energy cost of the upward reserve when reserves are sufficiently supplied.

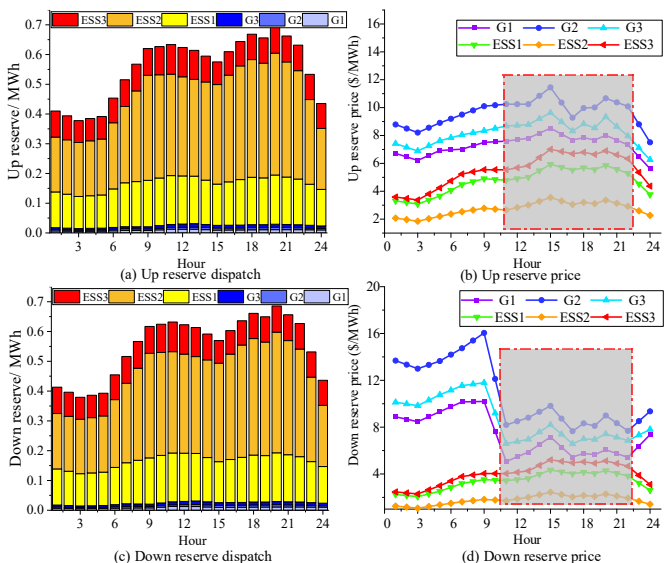

Fig. 6. Up/down reserve dispatch and local flexibility reserve price

Meanwhile, the reserve price difference for diverse providers reflects the true value of flexibility, as shown in *Proposition 3*. For example, the higher reserve price for G2 is caused by the distribution network congestion and unbalanced voltage constraints. Hence, reserves at neighboring nodes are largely offered by G2, while other flexibility reserve providers cannot satisfy the requirements. In this way, the diverse flexibility offered by various providers signifies the essence of network conditions in determining the values of flexibility reserves.

Fig. 7 compares maximum line loading in the deterministic case and risk-aware DRCC case, where the lack of coordination among local flexibility reserves in the deterministic case causes severe line congestion and hinders secure network operations.

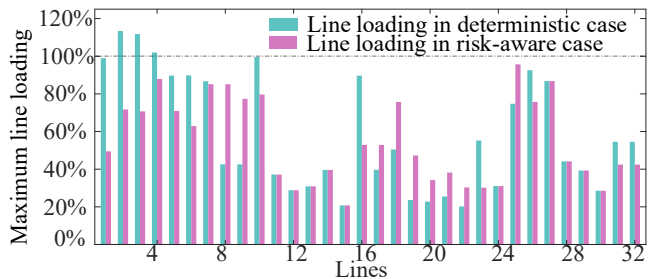

Fig. 7. Comparison of maximum line loading in deterministic and risk-aware DRCC cases

### C. Analysis of DSO's flexibility Pricing Scheme

The DSO's cleared flexibility prices are further analyzed to quantify the cost of managing the risk for spatially correlated uncertainties. Figs. 8 (a)-(b) present the cleared flexibility prices for charging uncertainties with respect to values of $\mu$ and $\sigma$. Sampling is performed every two hours.

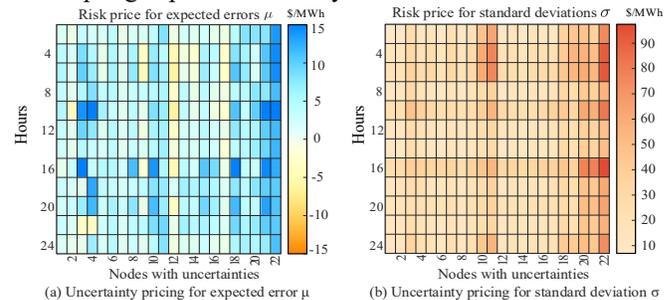

Fig. 8. DSO's flexibility pricing for expected errors $\mu$ and standard deviation $\sigma$ (sorted by the node number in ascending order)

Here, three-phase power distribution nodes with uncertainties includes those with distributed WTs or loads. A node with a high risk-aware price signal will pose a significant influence on the system-wide operating condition even with a small uncertainty fluctuation level. Generally, flexibility pricing for $\mu$ is less pricy than that of $\sigma$, indicating that the expected forecast error is relatively small with a minute impact on network operation. Fig. 8 depicts that the nodal flexibility pricing could vary within the same hour. In turn, the uncertainty risk has a dominant spatial effect as flexible resources are scattered throughout the network. Likewise, the hourly nodal operation varies according to the temporal fluctuation of uncertainty. The risk-aware flexibility pricing quantifies the uncertainty impact of nodal net injections on four parts, including energy, volt/var, and active /reactive power flows. The decomposition of price components reflects the network operating conditions and can guide flexible resources more intuitively.

Figs. 9 and 10 demonstrate that the system-wide operation risk is more sensitive to energy injection uncertainties at those nodes which possess higher prices for flexibility reserves. Typically, the energy component of flexibility price is relatively low for these nodes, which are located in the vicinity of flexible resources, e.g., 4th uncertain source (Node 8) installed with ESS, and 12th uncertain source (Node 19) with GT. In turn, flexible resources, such as distributed ESSs will be invested to be located in high uncertainty areas for improving the robustness of the distribution network operation.

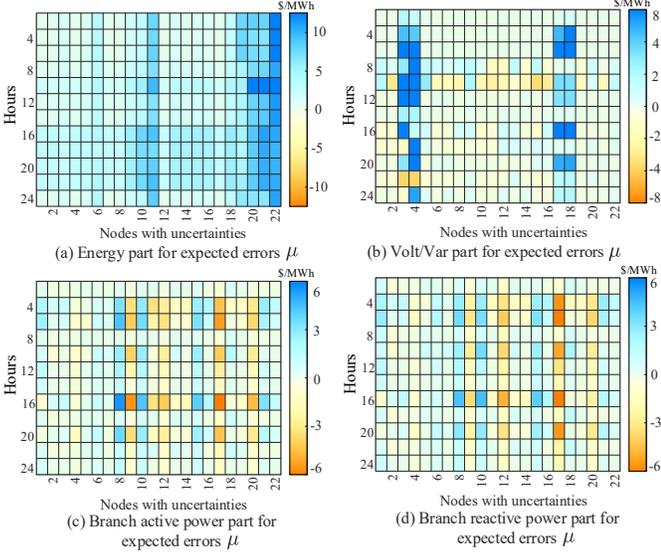

Fig. 9. Components of DSO's cleared flexibility pricing prices for $\mu$

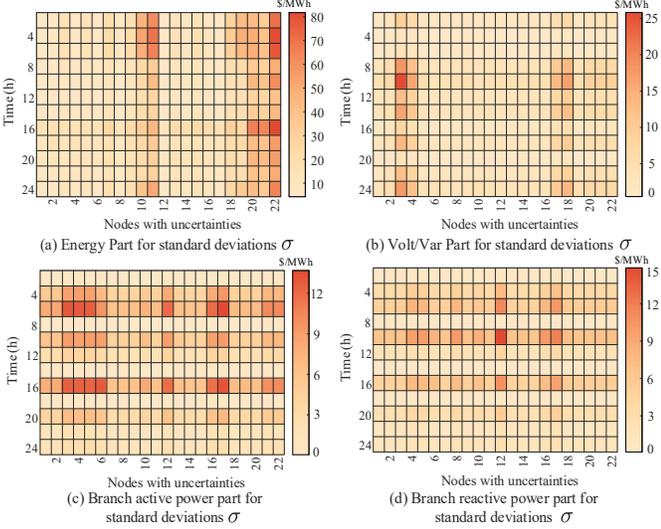

Fig. 10. Components of DSO's cleared flexibility prices for $\sigma$

Fig. 10 depicts that there are some differences in the volt/var part of the flexibility pricing scheme, which can be either positive or negative for $\mu$ and $\sigma$. This implies that the price component will be positive when the nodal voltage is lower than the reference voltage with a negative expected error μ, or the nodal voltage is higher than the reference voltage with a positive expected error μ. Otherwise, there will be a negative price component. That is, we can improve the distribution network performance by reducing the load value or increasing the generating power when the voltage is low and vice versa. In general, the volt/var price component values the impact of the uncertainties on the network voltage distribution. Figs. 9(b) and 10(b) depict that the absolute value of the price component for both flexibility prices provide a correlation with the degree of voltage deviation. The higher the nodal voltage deviation, the greater will be the impact of uncertainty on operating costs.

Furthermore, Figs. 9(c)-(d) and 10(c)-(d) show the price components on the power flow part for μ and σ, respectively. Once again, a larger absolute value indicates the nodal uncertainty will pose a greater network operation risk. Also, the absolute value of the power flow price tends to increase with the neighboring line loading, as shown in Fig. 7, such as 4th, 5th, 12th, and 17th uncertain sources (c.f. Nodes 8, 9, 19, and 26). It indicates when the line loading is high, the risk of line congestion will increase accordingly.

## VI. CONCLUSION

This paper proposes a risk-aware flexibility market clearing scheme for a DSO through a SDP-based probabilistic three-phase OPF solution. The solution quantifies the uncertainty risk and reward the use of local flexible resources in an unbalanced three-phase distribution network. In the proposed model, spatially correlated uncertainties are explicitly internalized into the DRCC probabilistic model through an information sharing mechanism. The paper derives the DSO's risk-aware flexible resource prices to mitigate the impact of network uncertainties on energy balance, voltage, and line losses, and network congestions. Case studies validate that the proposed analytic framework for the physical composition of risk-aware price scheme can provide a more intuitive guidance for the unbalanced three-phase distribution network operation. Meanwhile, resources are motivated to curb their variability by investing on local flexible resources or improving the quality of forecast information.

## APPENDIX

### A. Auxiliary variables for deriving a compact form of the unbalanced power flow

To perform a compact form of the unbalanced power flow, we introduce two auxiliary vectors $\boldsymbol{e}_i^\varphi$ and $\boldsymbol{h}_i^\varphi$ which are $(3i+\varphi)$-th standard basis vector in $\mathbb{R}^{3N}$ and $\mathbb{R}^{6N}$, where $\varphi=1, 2, 3$ represent phases a, b and c, respectively. Given the three-phase system admittance matrix as $\mathbf{Y}$, we use $\boldsymbol{y}_i^\varphi$ to describe a matrix with all zeros except for the $(3i+\varphi)$-th row, which corresponds to the $(3i+\varphi)$-th row of $\mathbf{Y}$. Based on $\boldsymbol{e}_i^\varphi$ and $\boldsymbol{h}_i^\varphi$, five auxiliary symmetric matrices $\boldsymbol{Y}_i^\varphi, \overline{\boldsymbol{Y}}_i^\varphi, \boldsymbol{M}_i^\varphi$ and $\boldsymbol{\Phi}_{i,j}^\phi, \overline{\boldsymbol{\Phi}}_{i,j}^\phi$ are further denoted.

$$\boldsymbol{y}_i^\varphi = \boldsymbol{e}_i^\varphi (\boldsymbol{e}_i^\varphi)^T \mathbf{Y} \quad \text{(A.1)}$$

$$\boldsymbol{Y}_i^\varphi = \frac{1}{2}\begin{bmatrix} \Re\left(\boldsymbol{y}_i^\varphi + (\boldsymbol{y}_i^\varphi)^T\right) & \Im\left((\boldsymbol{y}_i^\varphi)^T - \boldsymbol{y}_i^\varphi\right) \\ \Im\left(\boldsymbol{y}_i^\varphi - (\boldsymbol{y}_i^\varphi)^T\right) & \Re\left(\boldsymbol{y}_i^\varphi + (\boldsymbol{y}_i^\varphi)^T\right) \end{bmatrix} \quad \text{(A.2)}$$

$$\overline{\boldsymbol{Y}}_i^\varphi = -\frac{1}{2}\begin{bmatrix} \Im\left(\boldsymbol{y}_i^\varphi + (\boldsymbol{y}_i^\varphi)^T\right) & \Re\left(\boldsymbol{y}_i^\varphi - (\boldsymbol{y}_i^\varphi)^T\right) \\ \Re\left((\boldsymbol{y}_i^\varphi)^T - \boldsymbol{y}_i^\varphi\right) & \Im\left(\boldsymbol{y}_i^\varphi + (\boldsymbol{y}_i^\varphi)^T\right) \end{bmatrix} \quad \text{(A.3)}$$

$$\boldsymbol{M}_i^\varphi = \begin{bmatrix} \boldsymbol{e}_i^\varphi (\boldsymbol{e}_i^\varphi)^T & \boldsymbol{0} \\ \boldsymbol{0} & \boldsymbol{e}_i^\varphi (\boldsymbol{e}_i^\varphi)^T \end{bmatrix} \quad \text{(A.4)}$$

$$\boldsymbol{\Phi}_{i,j}^\varphi = \frac{1}{2}\cdot(\boldsymbol{e}_i^\varphi)^T \cdot \Re(\mathbf{Y}) \cdot \boldsymbol{e}_j^\varphi.$$

$$\begin{pmatrix} h_j^\phi (h_i^\phi)^T + h_i^\phi (h_j^\phi)^T + h_{j+n}^\phi (h_{i+n}^\phi)^T \\ + h_{i+n}^\phi (h_{j+n}^\phi)^T - 2 \cdot h_i^\phi (h_i^\phi)^T - 2 \cdot h_{i+n}^\phi (h_{i+n}^\phi)^T \end{pmatrix}$$
$$+ \frac{1}{2} \cdot (e_i^\varphi)^T \cdot \Im(Y) \cdot e_j^\varphi \cdot$$
$$\left( h_j^\varphi (h_{i+n}^\varphi)^T + h_{i+n}^\varphi (h_j^\varphi)^T - h_{j+n}^\varphi (h_i^\varphi)^T - h_i^\varphi (h_{j+n}^\varphi)^T \right) \quad \text{(A.5)}$$

$$\overline{\Phi}_{i,j}^\varphi = \frac{1}{2} \cdot (e_i^\varphi)^T \cdot \Im(Y) \cdot e_j^\varphi \cdot$$
$$\begin{pmatrix} 2 \cdot h_j^\varphi (h_i^\varphi)^T + 2 \cdot h_{i+n}^\varphi (h_{i+n}^\varphi)^T - h_j^\varphi (h_i^\varphi)^T \\ - h_i^\varphi (h_j^\varphi)^T - h_{j+n}^\varphi (h_{i+n}^\varphi)^T - h_{i+n}^\varphi (h_{j+n}^\varphi)^T \end{pmatrix}$$
$$- \frac{1}{2} \cdot (e_i^\varphi)^T \cdot \Re(Y) \cdot e_j^\varphi \cdot$$
$$\left( h_{j+n}^\varphi (h_i^\varphi)^T + h_i^\varphi (h_{j+n}^\varphi)^T - h_j^\varphi (h_{i+n}^\varphi)^T - h_{i+n}^\varphi (h_j^\varphi)^T \right) \quad \text{(A.6)}$$

where we use $\Re(A)$ and $\Im(A)$ to denote the real and the imaginary parts of matrix $A$. Superscript $T$ refers to the transpose of a matrix and $0$ denotes a matrix with all zeros.

### B. Detailed Modeling for Generator and ESSs Operational Limits:

➢ Generation and ramping limits:

$$g_i^{min} + R_{i,t}^{dn} \leq p_{i,t}^g \leq g_i^{max} - R_{i,t}^{up}, i \in \Omega_g: \underline{v}_{i,t}^{dn}, \overline{v}_{i,t}^{up} \quad \text{(B.1)}$$
$$-RD_i \cdot \Delta t \leq (p_{i,t}^g - p_{i,t-1}^g) \leq RU_i \cdot \Delta t, i \in \Omega_g: \underline{v}_{i,t}^{RU}, \overline{v}_{i,t}^{RD} \quad \text{(B.2)}$$
$$0 \leq R_{i,t}^{up} \leq RU_i \cdot \Delta t, i \in \Omega_g: \underline{v}_{i,t}^{RU}, \overline{v}_{i,t}^{RU} \quad \text{(B.3)}$$
$$0 \leq R_{i,t}^{dn} \leq RD_i \cdot \Delta t, i \in \Omega_g: \underline{v}_{i,t}^{RD}, \overline{v}_{i,t}^{RD} \quad \text{(B.4)}$$
$$p_{i,t}^g \cdot PF_i^{min} \leq q_{i,t}^g \leq p_{i,t}^g \cdot PF_i^{max}, i \in \Omega_g: \underline{v}_{i,t}^Q, \overline{v}_{i,t}^Q \quad \text{(B.5)}$$
$$q_{i,t}^g = \sum_\varphi q_{i,\varphi,t}^g, i \in \Omega_g: v_{i,t}^q \quad \text{(B.6)}$$
$$p_{i,t}^g = \sum_\varphi p_{i,\varphi,t}^g, i \in \Omega_g: v_{i,t}^p \quad \text{(B.7)}$$

where $p_{i,t}^g$, $q_{i,t}^g$ are active and reactive power generation at bus $i$, respectively. $RU_i/RD_i$ are up and down ramping rates. $PF_i^{min}/PF_i^{max}$ denote minimum/maximum power factors. The up/down reserves $R_{i,t}^{up}$, $R_{i,t}^{dn}$ are constrained by (B.1), (B.3), (B.4).

➢ ESSs charging/discharging limits:

$$ch_{i,t} = \sum_{\varphi \in \Phi_s} ch_{i,t}^\varphi, \; i \in \Omega_s: \mu_{i,t}^{ch} \quad \text{(B.8)}$$
$$dis_{i,t} = \sum_{\varphi \in \Phi_s} dis_{i,t}^\varphi, \; i \in \Omega_s: \mu_{i,t}^{dis} \quad \text{(B.9)}$$
$$q_{i,t}^s = \sum_{\varphi \in \Phi_s} q_{i,\varphi,t}^s, \; i \in \Omega_s: \mu_{i,t}^Q \quad \text{(B.10)}$$
$$SOC_{i,t} - SOC_{i,t-1} = \left( ch_{i,t} \cdot \eta_S - \frac{dis_{i,t}}{\eta_S} \right) \cdot \Delta t : \mu_{i,t}^{soc} \quad \text{(B.11)}$$
$$0 \leq SOC_{i,t} \leq SOC_i^{max}, i \in \Omega_s: \underline{\mu}_{i,t}^{soc}, \overline{\mu}_{i,t}^{soc} \quad \text{(B.12)}$$
$$ch_i^{min} \leq ch_{i,t} + R_{i,t}^{dn} \leq ch_i^{max}, i \in \Omega_s: \underline{\mu}_{i,t}^{ch}, \overline{\mu}_{i,t}^{ch} \quad \text{(B.13)}$$
$$dis_i^{min} \leq dis_{i,t} + R_{i,t}^{up} \leq dis_i^{max}, i \in \Omega_s: \underline{\mu}_{i,t}^{dis}, \overline{\mu}_{i,t}^{dis} \quad \text{(B.14)}$$
$$(dis_{i,t} - ch_{i,t})^2 + (q_{i,t}^s)^2 \leq (dis_i^{max})^2, i \in \Omega_s: \mu_{i,t}^s \quad \text{(B.15)}$$

where the left-hand side variables in (B.8)-(B.10) represent the sum of all variables over all phases. Given that SOC is the state of charge variable, $SOC_i^{max}$, $ch_i^{min}$, $ch_i^{max}$, $dis_i^{min}$, $dis_i^{max}$ represent the maximum SOC capacity, and lower/ upper limits of charging/discharging power, respectively. Besides providing active power support for the three-phase distribution network, ESSs provide reactive power support to a network, as shown in (B.10) and (B.15). The ESS reserves are constrained by (B.13)-(B.14).

### C. Proof for Proposition 1

Combining the sensitivity matrix (7h) with the nodal deviation matrix (9a), we obtain:

$$\Delta X = (JS)^{-1} \cdot \Delta S = (JS)^{-1} \cdot \begin{bmatrix} -A^\xi \cdot \xi + r \\ -\psi \odot A^\xi \cdot \xi + \psi \odot r \end{bmatrix}$$
$$= \begin{bmatrix} JP \\ JQ \end{bmatrix}^{-1} \cdot \begin{bmatrix} -A^\xi + \beta^\xi \\ -\psi \odot A^\xi + \psi \odot \beta^\xi \end{bmatrix} \cdot \xi \quad \text{(C.1)}$$

where $\psi$ denotes the vector of given power factors. Furthermore, based on (7c)-(7e) and (C.1), the following mapping relationships are derived:

$$\Delta |V_{i,\varphi,t}|^2 = JV_{i,t}^\varphi \cdot \begin{bmatrix} JP_t \\ JQ_t \end{bmatrix}^{-1} \cdot \begin{bmatrix} -A^\xi + \beta_t^\xi \\ -\psi \odot A^\xi + \psi \odot \beta_t^\xi \end{bmatrix} \cdot \xi_t \quad \text{(C.2)}$$

$$\Delta P_{i,j,t}^\varphi = JP_{i,j,t}^\varphi \cdot \begin{bmatrix} JP_t \\ JQ_t \end{bmatrix}^{-1} \cdot \begin{bmatrix} -A^\xi + \beta_t^\xi \\ -\psi \odot A^\xi + \psi \odot \beta_t^\xi \end{bmatrix} \cdot \xi_t \quad \text{(C.3)}$$

$$\Delta Q_{i,j,t}^\varphi = JQ_{ij,t}^\varphi \cdot \begin{bmatrix} JP_t \\ JQ_t \end{bmatrix}^{-1} \cdot \begin{bmatrix} -A^\xi + \beta_t^\xi \\ -\psi \odot A^\xi + \psi \odot \beta_t^\xi \end{bmatrix} \cdot \xi_t \quad \text{(C.4)}$$

Based on the Taylor's theorem, we can further obtain the system-wide response functions of voltages and active and reactive power flows to uncertainties as shown in (10a)-(10c)

### D. Proof for Proposition 3

This section provides the proof for *Proposition 3*. The flexibility reserve prices can be obtained via taking the first order partial derivatives of $\beta_{i,k,t}^\varphi$ to zero.

$$\frac{\partial \mathcal{L}}{\partial \beta_{i,k,t}^\varphi} = -\lambda_{t,k}^R + \frac{z_R \cdot (e_k)^T \cdot r_t \cdot \beta_{i,t}^\varphi}{\|r_t^{1/2} \cdot \beta_{i,t}^\varphi\|_2} \cdot (\overline{\alpha}_{i,t}^R + \underline{\alpha}_{i,t}^R)$$
$$+ (\overline{\alpha}_{i,t}^R - \underline{\alpha}_{i,t}^R) \cdot \mu_{k,t}$$
$$+ \mu_{k,t} \cdot \sum_{i,\varphi} (\overline{\alpha}_{i,\varphi,t}^V - \underline{\alpha}_{i,\varphi,t}^V) \cdot JV_{i,t}^\varphi \cdot \begin{bmatrix} JP_t \\ JQ_t \end{bmatrix}^{-1} \cdot \begin{bmatrix} e_i^\varphi \\ \psi \odot e_i^\varphi \end{bmatrix}$$
$$+ \sum_{i,\varphi} (\overline{\alpha}_{i,\varphi,t}^V + \underline{\alpha}_{i,\varphi,t}^V) \cdot Z_{V_{i,t}^\varphi}$$
$$+ \mu_{k,t} \cdot \sum_{ij \in L,\varphi} (\overline{\alpha}_{ij,t}^p - \underline{\alpha}_{ij,t}^p) \cdot JP_{ij,t}^\varphi \cdot \begin{bmatrix} JP_t \\ JQ_t \end{bmatrix}^{-1} \cdot \begin{bmatrix} e_i^\varphi \\ \psi \odot e_i^\varphi \end{bmatrix}$$
$$+ \sum_{ij \in L} (\overline{\alpha}_{ij,t}^p + \underline{\alpha}_{ij,t}^p) \cdot Z_{P_{ij,t}}$$
$$+ \mu_{k,t} \cdot \sum_{ij \in L,\varphi} (\overline{\alpha}_{ij,t}^q - \underline{\alpha}_{ij,t}^q) \cdot JQ_{ij,t}^\varphi \cdot \begin{bmatrix} JP_t \\ JQ_t \end{bmatrix}^{-1} \cdot \begin{bmatrix} e_i^\varphi \\ \psi \odot e_i^\varphi \end{bmatrix}$$
$$+ \sum_{ij \in L} (\overline{\alpha}_{ij,t}^p + \underline{\alpha}_{ij,t}^p) \cdot Z_{Q_{ij,t}} = 0. \quad \text{(D.1)}$$

The above derivation can be reformulated as below:

$$\left( \frac{z_R \cdot (e_k)^T \cdot r_t \cdot \beta_{i,t}^\varphi}{\|r_t^{1/2} \cdot \beta_{i,t}^\varphi\|_2} + \mu_{k,t} \right) \cdot \overline{\alpha}_{i,t}^R + \left( \frac{z_R \cdot (e_k)^T \cdot r_t \cdot \beta_{i,t}^\varphi}{\|r_t^{1/2} \cdot \beta_{i,t}^\varphi\|_2} + \mu_{k,t} \right) \cdot \underline{\alpha}_{i,t}^R$$
$$= \lambda_{t,k}^R - \sum_{i,\varphi} (\overline{\alpha}_{i,\varphi,t}^V + \underline{\alpha}_{i,\varphi,t}^V) \cdot Z_{V_{i,t}^\varphi}$$
$$- \sum_{ij \in L} (\overline{\alpha}_{ij,t}^p + \underline{\alpha}_{ij,t}^p) \cdot Z_{P_{ij,t}} - \sum_{ij \in L} (\overline{\alpha}_{ij,t}^p + \underline{\alpha}_{ij,t}^p) \cdot Z_{Q_{ij,t}}$$
$$- \mu_{k,t} \cdot \sum_{i,\varphi} (\overline{\alpha}_{i,\varphi,t}^V - \underline{\alpha}_{i,\varphi,t}^V) \cdot JV_{i,t}^\varphi \cdot \begin{bmatrix} JP_t \\ JQ_t \end{bmatrix}^{-1} \cdot \begin{bmatrix} e_i^\varphi \\ \psi \odot e_i^\varphi \end{bmatrix}$$
$$- \mu_{k,t} \cdot \sum_{ij \in L,\varphi} (\overline{\alpha}_{ij,t}^p - \underline{\alpha}_{ij,t}^p) \cdot JP_{ij,t}^\varphi \cdot \begin{bmatrix} JP_t \\ JQ_t \end{bmatrix}^{-1} \cdot \begin{bmatrix} e_i^\varphi \\ \psi \odot e_i^\varphi \end{bmatrix}$$
$$- \mu_{k,t} \cdot \sum_{ij \in L,\varphi} (\overline{\alpha}_{ij,t}^q - \underline{\alpha}_{ij,t}^q) JQ_{ij,t}^\varphi \begin{bmatrix} JP_t \\ JQ_t \end{bmatrix}^{-1} \begin{bmatrix} e_i^\varphi \\ \psi \odot e_i^\varphi \end{bmatrix} \quad \text{(D.2)}$$

When $\mu_{k,t} = 0$, (D.2) will be simplified as Eq. (13a). Herein, *Proposition 3* is proven.

*E. Proof for Proposition 5 (Profit sufficiency)*

This section provides the proof for *Proposition 5*. Based on (12a)-(12b) and the corresponding complementary slackness, we can have the following relationship:

$$C_{i,t}^R := \bar{\alpha}_{i,t}^R \cdot R_{i,t}^{\text{up}} + \underline{\alpha}_{i,t}^R \cdot R_{i,t}^{\text{dn}}$$

$$= \bar{\alpha}_{i,t}^R \cdot \left(\sum_\varphi (\boldsymbol{\beta}_{i,t}^\varphi)^T \cdot \boldsymbol{\mu}_t + z_R \|\boldsymbol{\Gamma}_t^{1/2} \cdot \boldsymbol{\beta}_{i,t}^\varphi\|_2\right)$$

$$+ \underline{\alpha}_{i,t}^R \cdot \left(-z_R \|\boldsymbol{\Gamma}_t^{1/2} \cdot \boldsymbol{\beta}_{i,t}^\varphi\|_2 + \sum_\varphi (\boldsymbol{\beta}_{i,t}^\varphi)^T \cdot \boldsymbol{\mu}_t\right)$$

$$= \sum_{\varphi,k} \bar{\alpha}_{i,t}^R \cdot \left(\mu_k + \frac{z_R \cdot (e_k)^T \cdot \boldsymbol{\Gamma}_t \cdot \boldsymbol{\beta}_{i,t}^\varphi}{\|\boldsymbol{\Gamma}_t^{1/2} \cdot \boldsymbol{\beta}_{i,t}^\varphi\|_2}\right) \cdot \beta_{i,k,t}^\varphi$$

$$+ \sum_{\varphi,k} \underline{\alpha}_{i,t}^R \cdot \left(-\mu_k + \frac{z_R \cdot (e_k)^T \cdot \boldsymbol{\Gamma}_t \cdot \boldsymbol{\beta}_{i,t}^\varphi}{\|\boldsymbol{\Gamma}_t^{1/2} \cdot \boldsymbol{\beta}_{i,t}^\varphi\|_2}\right) \cdot \beta_{i,k,t}^\varphi$$

$$= \sum_{\varphi,k} \beta_{i,k,t}^\varphi \cdot \begin{pmatrix} (\bar{\alpha}_{i,t}^R + \underline{\alpha}_{i,t}^R) \cdot \frac{z_R \cdot (e_k)^T \cdot \boldsymbol{\Gamma}_t \cdot \beta_{i,k,t}^\varphi}{\|\boldsymbol{\Gamma}_t^{1/2} \cdot \boldsymbol{\beta}_{i,t}^\varphi\|_2} \\ + (\bar{\alpha}_{i,t}^R - \underline{\alpha}_{i,t}^R) \cdot \mu_{k,t} \end{pmatrix} \quad \text{(E.1)}$$

Similarly, the uncertainty payment is calculated as:

$$E_{k,t}^\xi := \lambda_{\mu,k,t} \cdot \mu_k + \lambda_{\sigma,k,t} \cdot \sigma_{k,t}$$

$$= \sum_{i,\varphi} \beta_{i,k,t}^\varphi \cdot \begin{bmatrix} (\bar{\alpha}_{i,t}^R - \underline{\alpha}_{i,t}^R) \cdot \mu_{k,t} + \\ (\bar{\alpha}_{i,t}^R + \underline{\alpha}_{i,t}^R) \cdot \frac{z_R \cdot (e_k)^T \cdot \boldsymbol{\Gamma}_t \cdot \boldsymbol{\beta}_{i,t}^\varphi}{\|\boldsymbol{\Gamma}_t^{1/2} \cdot \boldsymbol{\beta}_{i,t}^\varphi\|_2} \end{bmatrix}$$

$$+ \sum_{i,\varphi} XV_{i,k,t}^\varphi \cdot \begin{bmatrix} (\bar{\alpha}_{i,\varphi,t}^V - \underline{\alpha}_{i,\varphi,t}^V) \cdot \mu_{k,t} + \\ (\bar{\alpha}_{i,\varphi,t}^V + \underline{\alpha}_{i,\varphi,t}^V) \cdot \frac{z_v \cdot XV_{i,t}^\varphi \cdot \boldsymbol{\Gamma}_t \cdot e_k}{\|XV_{i,t}^\varphi \cdot \boldsymbol{\Gamma}_t^{1/2}\|_2} \end{bmatrix}$$

$$+ \sum_{l,\varphi} XP_{l,k,t}^\varphi \cdot \begin{bmatrix} (\bar{\alpha}_{l,t}^p - \underline{\alpha}_{l,t}^p) \cdot \mu_{k,t} + \\ (\bar{\alpha}_{l,t}^p + \underline{\alpha}_{l,t}^p) \cdot \frac{z_l \cdot \sum_\varphi XP_{l,t}^\varphi \cdot \boldsymbol{\Gamma}_t \cdot e_k}{\|\sum_\varphi XP_{l,t}^\varphi \cdot \boldsymbol{\Gamma}_t^{1/2}\|_2} \end{bmatrix}$$

$$+ \sum_{l,\varphi} XQ_{l,k,t}^\varphi \cdot \begin{bmatrix} (\bar{\alpha}_{l,t}^q - \underline{\alpha}_{l,t}^q) \cdot \mu_{k,t} + \\ (\bar{\alpha}_{l,t}^q + \underline{\alpha}_{l,t}^q) \cdot \frac{z_l \cdot \sum_\varphi XQ_{l,t}^\varphi \cdot \boldsymbol{\Gamma}_t \cdot e_k}{\|\sum_\varphi XQ_{l,t}^\varphi \cdot \boldsymbol{\Gamma}_t^{1/2}\|_2} \end{bmatrix} \quad \text{(E.2)}$$

by comparing $\sum_i C_{i,t}^R$ and $\sum_k E_{k,t}^\xi$, the payment $\sum_k E_{k,t}^\xi$ can be formulated as the reserve cost plus flexibility cost for the distribution network operation margin, including the flexibility margins for voltage and active and reactive power flows.

$$\sum_k E_{k,t}^\xi = \sum_i C_{i,t}^R + \sum_k C_{k,t}^V + \sum_k C_{k,t}^P + \sum_k C_{k,t}^Q \quad \text{(E.3)}$$